\documentclass[preprint,noshowkeys,amsmath,amssymb,superscriptaddress,nofootinbib,mathtools]{revtex4-1}

\usepackage{bm}
\usepackage{slashed}
\usepackage{ulem}

\usepackage[pdftex]{graphicx,color}
\usepackage{epstopdf}
\usepackage[bookmarks=true,bookmarksnumbered=true,bookmarkstype=toc, colorlinks=true, citecolor=blue, linkcolor=blue ,urlcolor=blue]{hyperref} 
\usepackage{units}

\newcommand{\lsim}{\raise0.3ex\hbox{$\;<$\kern-0.75em\raise-1.1ex\hbox{$\sim\;$}}}
\newcommand{\gsim}{\raise0.3ex\hbox{$\;>$\kern-0.75em\raise-1.1ex\hbox{$\sim\;$}}}
\usepackage{mathtools}

\renewcommand{\thesubsection}{\arabic{section}.\arabic{subsection}}

\numberwithin{equation}{section}

\renewcommand{\eqref}[1]{Eq.~(\ref{#1})}

\usepackage{color}

\definecolor{fgreen}{cmyk}{0.91,0,0.88,0.12}

\definecolor{pink}{cmyk}{0,0.5,0,0}

\definecolor{pastelpink}{cmyk}{0,0.25,0,0}

\definecolor{softpink}{cmyk}{0,0.125,0,0}

\definecolor{purple}{cmyk}{0.5,1.0,0.1,0}

\definecolor{violet}{cmyk}{0.75,1,0.25,0}

\usepackage{ulem}  

\usepackage[markup=underlined,addedmarkup=colored,defaultcolor=fgreen]{changes}

\newcommand{\lmlt}{L_\mu - L_\tau}

\usepackage[utf8]{inputenc}

\begin{document}

\preprint{UME-PP-023}
\preprint{KYUSHU-HET-247}
\preprint{KYUSHU-RCAPP-2022-04}

\title{Search for Lepton Flavor Violating Decay at FASER}

\author{Takeshi Araki}
\email{t-araki@den.ohu-u.ac.jp}
\affiliation{Faculty of Dentistry, Ohu University, 31-1 Sankakudo, Tomita-machi, Koriyama, Fukushima 963-8611,  Japan}

\author{Kento Asai}
\email{kento@icrr.u-tokyo.ac.jp}
\affiliation{Institute for Cosmic Ray Research (ICRR), The University of Tokyo, Kashiwa, Chiba 277--8582, Japan}
\affiliation{Department of Physics, Faculty of Engineering Science, Yokohama National University, Hodogaya-ku, Yokohama 240-8501, Japan}
\affiliation{Department of Physics, Faculty of Science, Saitama University, Sakura-ku, Saitama 338--8570, Japan}

\author{Hidetoshi Otono}
\email{otono@phys.kyushu-u.ac.jp}
\affiliation{Research Center for Advanced Particle Physics, Kyushu University, 744 Motooka, Nishi-ku, Fukuoka 819-0395, Japan}

\author{Takashi Shimomura}
\email{shimomura@cc.miyazaki-u.ac.jp}
\affiliation{Faculty of Education, University of Miyazaki, 1-1 Gakuen-Kibanadai-Nishi, Miyazaki 889-2192, Japan}
\affiliation{Department of Physics, Kyushu University, 744 Motooka, Nishi-ku, Fukuoka, 819–0395, Japan}

\author{Yosuke Takubo}
\email{yosuke.takubo@kek.jp}
\affiliation{Institute of Particle and Nuclear Studies, High Energy Accelerator Research Organization (KEK), Oho 1-1, Tsukuba, Ibaraki 305-0801, Japan}

\begin{abstract}
FASER is one of the promising experiments which search for long-lived particles beyond the Standard Model.
In this paper, we consider charged lepton flavor violation (CLFV) via a light and weakly interacting boson and discuss the detectability by FASER.
We focus on four types of CLFV interactions, i.e., the scalar-, pseudoscalar-, vector-, and dipole-type interaction, and calculate the sensitivity of FASER to each CLFV interaction.
We show that, with the setup of FASER2, a wide region of the parameter space can be explored.
Particularly, it is found that FASER2 has a sensitivity to very small coupling regions in which the rare muon decays, such as $\mu \rightarrow e\gamma$, cannot place bounds, and that there is a possibility to detect CLFV decays of the new light bosons.
\end{abstract}


\maketitle

\section{Introduction}
\label{sec:introduction}
The discovery of neutrino oscillation has revealed that lepton flavor is not conserved in nature. 
The non-conservation of lepton flavor is expected to occur in the charged lepton sector as well as the neutral lepton sector. 
Searches for the charged lepton flavor violation (CLFV) have been performed over decades in rare processes, 
such as radiative decays and muon-electron conversions in nuclei.
Other searches have been performed in collider experiments seeking CLFV decays of hypothetical heavy new particles.
However, neither such CLFV rare processes nor CLFV decays of heavy particles have been observed yet.  
Null results of these searches motivate us to consider other possibilities of CLFV processes with
light and feebly interacting new particles. 

CLFV interactions are generally predicted in models with the generation of neutrino masses and mixing. 
One of the most well-studied models in this regard is the one with a new scalar boson responsible for neutrino masses. 
Such scalar-type CLFV interactions originate from Yukawa interactions with the new scalar bosons in extensions of the 
Standard Model (SM), such as 
two Higgs doublet models \cite{Kubo:2006yx}
and type-II seesaw models \cite{Ma:2001mr,Chun:2003ej,Kakizaki:2003jk,Abada:2007ux,Akeroyd:2009nu,Fukuyama:2009xk,Fukuyama:2010mz,Primulando:2019evb}.
After diagonalizing a mass matrix of the charged leptons, the misalignment of Yukawa couplings between the SM Higgs boson and 
the new scalar boson results in the CLFV interactions. 
Axion-like-particle (ALP) type interaction is another interesting case of scalar interactions. 
The ALP is a pseudo Nambu-Goldstone boson emerging from the spontaneous breaking of a global symmetry, which 
has been discussed in context of the strong CP problem~\cite{Peccei:1977hh,Peccei:1977ur,Weinberg:1977ma,Wilczek:1977pj}, the anomalies in $B$ meson decays~\cite{Bhattacharya:2021shk,Bauer:2021mvw,Bonilla:2022qgm} and the origin of dark matter~\cite{Preskill:1982cy,Abbott:1982af,Dine:1982ah}. 
CLFV interactions of the ALP have been studied in Refs.~\cite{Heeck:2017xmg, Cornella:2019uxs,Endo:2020mev,Calibbi:2020jvd}.
Another type of CLFV interactions is the one with a new gauge boson. Vector-type CLFV interactions 
can appear in flavor or family gauge symmetric models, such as gauged U(1)$_{L_\alpha - L_\beta}$ models 
($\alpha,\beta= e,\mu,\tau$)~\cite{Foot:1990mn,He:1990pn,He:1991qd,Foot:1994vd}.
When the charged leptons are non-universally charged under the new gauge symmetry and have non-diagonal Yukawa couplings, the CLFV interactions emerge in the gauge sector after the symmetry breaking \cite{Foot:1994vd,Heeck:2016xkh,Altmannshofer:2016brv,Iguro:2020rby,Cheng:2021okr}.
Furthermore, when we consider nonrenormalizable dimension $5$ operators, the dipole-type CLFV interactions are possible even 
in flavor universal gauge symmetric models, such as gauged U(1)$_{B-L}$ models and dark photon models. 
Such dipole-type CLFV interactions are generated at loop level by integrating 
out heavy scalars and/or fermions propagating in the loop \cite{Nomura:2020azp}. 
The CLFV decays through these interactions have been extensively studied in various models and 
searched in experiments, 
assuming that the new scalar or gauge bosons are much heavier than the electroweak (EW) breaking scale.

Although heavy new particles are generally considered in the literature, the new scalar or gauge 
bosons can be lighter than the EW symmetry breaking scale, if their interactions with the SM particles are feeble.  
Recently, such light and feebly interacting particles have been receiving attention in various fields: 
the muon $g-2$ anomaly \cite{Foot:1994vd,Pospelov:2008zw,Altmannshofer:2014pba,Bodas:2021fsy,Ghorbani:2021yiw}, 
rare decays of mesons \cite{Altmannshofer:2014cfa}, and 
the observations of high energy cosmic neutrinos \cite{Ioka:2014kca,Ng:2014pca,Ibe:2014pja,Araki:2014ona,DiFranzo:2015qea,Araki:2015mya,Bustamante:2020mep,Carpio:2021jhu}. 
Due to the feeble interactions, the new bosons are expected to be long-lived and can travel more than a hundred meters 
from their production points.

FASER (ForwArd Search ExpeRiment) \cite{Feng:2017uoz,Feng:2017vli,Kling:2018wct,Feng:2018pew,FASER:2018ceo,Ariga:2018uku,Ariga:2018pin,Ariga:2019ufm} is a new experiment to search for such new light, feebly interacting, neutral particles, that are generated from proton-proton collisions in the Large Hadron Collider (LHC) of the European Organization for Nuclear Research (CERN). 
The FASER detector is placed 480 m downstream from the ATLAS interaction point (IP) and detects charged particles from new particle decays.
FASER started physics data-taking in July 2022 and will collect more than $150$ fb$^{-1}$ of data during the LHC Run $3$ (2022-2025). The upgrade of the FASER detector to take $3$ ab$^{-1}$ at the High-Luminosity LHC (FASER2) is also being discussed. 

Due to the high luminosity of the proton-proton interactions at the forward region, FASER will realize high sensitivity to not only charged lepton flavor conserving (CLFC) decays but also CLFV decays of the new bosons.
In this paper, we consider the CLFV decays of the light and long-lived new bosons for the scalar-, pseudoscalar-, vector-, and dipole-type interactions, and discuss the sensitivity to CLFV couplings of FASER. 
For analysis of the new gauge boson, we take into account the production from a new Higgs boson which gives the origin of the gauge boson mass and decays into a pair of the new gauge bosons. 
This new production process was recently studied in Ref.~\cite{Araki:2020wkq}.

This paper is organized as follows. In section \ref{sec:Lagrangian}, we present the interaction 
Lagrangians for the scalar-, pseudoscalar-, vector-, and dipole-type CLFV interactions.
The detectability of CLFV decays at FASER is briefly discussed in section \ref{sec:prod-detect},  
and formulae for the expected number of signal events are shown in section \ref{sec:production}.   
Our results for the sensitivity to the CLFV couplings are shown in section \ref{sec:result}. 
Section \ref{sec:summary} is devoted to summary and discussion.
In appendices \ref{apdx:decay-position} and \ref{apdx:decay-probability}, we give details of the event calculations.

\section{Interaction Lagrangian}
\label{sec:Lagrangian}
We study CLFV decays of light bosons for four types of interactions which we refer to as the scalar-, the pseudoscalar-, the vector-, and the dipole-type interaction. 
As we discuss later, the FASER detector will be able to identify electrons and muons, whereas the identification of tau leptons is difficult. 
Therefore, we introduce CLFV couplings only in the electron-muon ($e\mu$) sector.
Some of the interaction Lagrangians shown in this section were recently studied in Ref.~\cite{Araki:2021vhy}, in which their possible origins were also discussed based on multi Higgs doublet models, ALP models, gauged U(1)$_{\lmlt}$ models, and loop-induced dark photon models. 
Thus, we omit the details of their origins and only show the relevant Lagrangians to our analyses.

\subsection{Scalar-type interaction}
For the scalar-type interaction, we introduce a new scalar boson  $\phi_l$ which interacts with the SM particles through mixing with the SM Higgs boson $h$. 
In addition, we introduce Yukawa-type CLFV interactions of $\phi_l$ in the $e\mu$ sector. 
Then, the interaction Lagrangian of $\phi_l$ to the SM fermions is given by 
\begin{align}
\mathcal{L}_{\rm scalar} = 
\frac{\theta_{h\phi}}{v} \sum_{f} m_f \overline{f} \phi_l f
+\left( 
 y_{e\mu}\overline{e_L} \phi_l \mu_R
+y_{\mu e}\overline{\mu_L} \phi_l e_R
+ {\rm H.c.}
\right)~,
\label{eq:Lscl}
\end{align}
where $\theta_{h\phi}$ is the mixing angle between $\phi_l$ and $h$, and CLFV coupling constants are denoted as $y_{e \mu}$ and $y_{\mu e}$. 
The symbol $f$ in the first term runs over all the SM fermions with the mass $m_f$, 
and $L$ and $R$ denote left-handed and right-handed chirality.
The vacuum expectation value (VEV) of the SM Higgs boson is defined as $v=246$ GeV.

With \eqref{eq:Lscl}, the total decay width of $\phi_l$ is given by 
\begin{align}
    \Gamma_{\mathrm{total}} = 
      \Gamma(\phi_l \rightarrow {\rm hadrons})
      + \sum_{\ell = e,\mu,\tau}\Gamma(\phi_l \rightarrow \ell\bar{\ell})
      + \Gamma(\phi_l \rightarrow e\bar{\mu})
      + \Gamma(\phi_l \rightarrow \mu\bar{e})~.
\end{align}
The first term represents partial decay widths into all possible hadronic final states, and we use the decay widths provided in Ref.~\cite{Kling:2021fwx}.
The partial decay width into the charged leptons $\ell$ and $\ell^\prime$ is written as
\begin{align}
\Gamma(\phi_l \rightarrow \ell\bar{\ell'})
= \frac{1}{16\pi}m_\phi~
\lambda\left( \frac{m_\ell^2}{m_\phi^2}, \frac{m_{\ell'}^2}{m_\phi^2} \right)
\left[ 
S_1 \left( 1-\frac{m_\ell^2+m_{\ell'}^2}{m_\phi^2} \right)
-4 S_2 \frac{m_\ell m_{\ell'}}{m_\phi^2}
\right]~,
\end{align}
where $m_\phi$ and $m_{\ell(\ell')}$ stand for the masses of $\phi_l$ and $\ell(\ell')$, respectively, and the function $\lambda$ is the Kallen function defined as follows~:
\begin{align}
\lambda(a,b) = \sqrt{1 + a^2 + b^2 - 2a - 2b - 2 ab}~.
\label{eq:kallen}
\end{align}
The constants $S_1$ and $S_2$ are defined as $S_1 = 2 S_2= 2(\theta_{h\phi} m_\ell)^2/v^2$ for CLFC decays, while $S_1 = |y_{e\mu}|^2 + |y_{\mu e}|^2$ and $S_2 = {\rm Re}(y_{e\mu} y_{\mu e})$ for CLFV decays.

\subsection{Pseudoscalar interaction}
For the pseudoscalar-type interaction, we add an ALP $a$ to the SM particle content and introduce CLFV couplings in the $e\mu$ sector.
The relevant Lagrangian is given by 
\begin{align}
\label{eq:Lalp}
    \mathcal{L}_\textrm{pseudoscalar} =
    \frac{\partial_\rho a}{\Lambda} \left\{ \sum_{f} c_{ff} \overline{f} \gamma^\rho \gamma_5 f + c_{e\mu} \overline{e} \gamma^\rho \gamma_5 \mu + c_{e \mu}^\ast \overline{\mu} \gamma^\rho \gamma_5 e \right\}~,
\end{align}
where $c_{ff}$ and $c_{e \mu}$ are CLFC and CLFV coupling constants, respectively, and $\Lambda$ is a cutoff scale. 

From \eqref{eq:Lalp}, the total decay width of $a$ is given by~\footnote{
According to Refs.~\cite{Dolan:2014ska,Domingo:2016yih,Ariga:2018uku}, decays of the ALPs into light hadrons are forbidden or suppressed by the CP invariance and flavor universality of the coupling constants.
Therefore, we consider decays of the ALP only into $e^+e^-, \mu^+\mu^-, \tau^+\tau^-, \bar{c}c$, and $\bar{b}b$ in this paper.
}
\begin{align}
    \Gamma_{\mathrm{total}} =
    \Gamma(a \to {\rm hadrons}) + \sum_{\ell} \Gamma(a \to \ell \overline{\ell})
    + \Gamma(a \to e\overline{\mu}) + \Gamma(a \to \mu \overline{e})~,
\end{align}
where the partial decay width into charged leptons are given by~\cite{Heeck:2017xmg,Calibbi:2020jvd}
\begin{align}
\Gamma(a \to \ell \bar{\ell'})
= \frac{|c_{\ell \ell'}|^2}{8\pi \Lambda^2} m_a (m_\ell+m_{\ell'})^2
\lambda\left( \frac{m_\ell^2}{m_a^2}, \frac{m_{\ell'}^2}{m_a^2}\right)
\left[ 1 - \frac{(m_\ell - m_{\ell'})^2}{m_a^2} \right]~,
\end{align}
and $m_a$ stands for the ALP mass.

\subsection{Vector-type interaction}
For the vector-type interaction, we follow the discussion given in Ref.~\cite{Araki:2021vhy} and consider a broken gauged U(1)$_{\lmlt}$ model.
The CLFV interactions are parameterized by a mixing angle between an electron and a muon, and we denote it as $\theta_{\mathrm{LFV}}$. 
The mass and flavor eigenstate of charged leptons are connected by this 
mixing angle.
Then, the Lagrangian of the vector-type interaction is given by
\begin{align}
\mathcal{L}_{\mathrm{vector}} &= g_{Z'} Z'_\rho (
s^2~ \overline{e} \gamma^\rho e 
+ c^2~ \overline{\mu} \gamma^\rho \mu
+ sc~ \overline{\mu} \gamma^\rho e 
+ sc~ \overline{e} \gamma^\rho \mu) \nonumber \\
&\qquad +g_{Z'} Z'_\rho (- \overline{\tau} \gamma^\rho \tau
+ \overline{\nu_\mu} \gamma^\rho \nu_\mu
- \overline{\nu_\tau} \gamma^\rho \nu_\tau
)~,
\label{eq:Lvec}
\end{align}
where $Z'$ and $g_{Z'}$ are the new gauge boson and its gauge coupling constant, respectively, while $s=\sin\theta_{\rm LFV}$ and $c=\cos\theta_{\rm LFV}$. 
Here, $\nu_\mu$ and $\nu_\tau$ are left-handed muon and tau neutrinos. 
For simplicity, we omit the kinetic mixing throughout this paper by assuming it is negligibly small.
In \eqref{eq:Lvec}, the U(1)$_{\lmlt}$ gauge symmetry is restored to the gauge interaction in the limit of $\theta_{\rm LFV} \to 0$.

Given the Lagrangian in \eqref{eq:Lvec}, the total decay width of $Z'$ is obtained as
\begin{align}
\Gamma_{\mathrm{total}} = 
\Gamma(Z' \rightarrow \nu\bar{\nu}) 
+ \sum_{\ell=e, \mu,\tau}\Gamma(Z' \rightarrow \ell\bar{\ell})
+ \Gamma(Z' \rightarrow e\bar{\mu})
+ \Gamma(Z' \rightarrow \mu\bar{e})~,
\end{align}
where the partial decay width into a neutrino pair is given by
\begin{align}
\Gamma(Z' \rightarrow \nu\bar{\nu}) = \frac{g_{Z'}^2}{12\pi}m_{Z'}~,
\end{align}
in the massless limit of neutrinos~\footnote{Here, we assumed neutrinos are Dirac particles. For Majorana neutrinos, the partial decay width is multiplied by $1/2$.}, and $m_{Z'}$ stands for the mass of $Z'$.
The partial decay width into charged leptons is written as 
\begin{align}
\Gamma(Z' \rightarrow \ell\bar{\ell'}) 
&= \frac{V^2}{24 \pi} m_{Z'}~\lambda\left(
1, \frac{m_\ell^2}{m_{Z'}^2}, \frac{m_{\ell'}^2}{m_{Z'}^2}\right) \nonumber \\
&\quad \times \left[
2
- \frac{m_\ell^2 - 6 m_\ell m_{\ell'} + m_{\ell'}^2}{m_{Z'}^2}
- \frac{(m_\ell^2 - m_{\ell'}^2)^2}{m_{Z'}^4}
\right]~,
\end{align}
where $V=g_{Z'} s^2$ or $g_{Z'} c^2$ for CLFC decays into $ee$ or $\mu\mu$, respectively, while $V=g_{Z'} s c$ for CLFV decays.

\subsection{Dipole-type interaction}
For the dipole-type interaction, we introduce a new U(1) gauge symmetry similarly to the vector-type interaction.
We, however, assume that all the SM particles are uncharged under the new U(1) gauge symmetry, and the new gauge boson has no interactions with the SM particles at tree level.
Even in such a case, interactions between the new gauge boson and the SM fermions can be induced at loop level if there are additional particles connecting the new gauge boson to the SM fermions.
In this work, we consider the following dipole interactions between the new gauge boson, $A'$, and the SM charged leptons: 
\begin{align}
    \mathcal{L}_{\mathrm{dipole}} &=
    \frac{1}{2} \sum_{\ell=e,\mu,\tau} \mu_\ell \overline{\ell} \sigma^{\rho \sigma} \ell A'_{\rho \sigma}
    + \frac{\mu'}{2} 
      \left( 
          \overline{\mu} \sigma^{\rho \sigma}e 
        + \overline{e} \sigma^{\rho \sigma}\mu 
      \right) A'_{\rho \sigma}~,
\label{eq:Ldpl}
\end{align}
where $\mu^\prime$ and $\mu_\ell$ are CLFV and CLFC dipole couplings, respectively, and $A_{\rho\sigma}^\prime$ stands for the field strength of $A^\prime$. 
Here, the dipole couplings are assumed to be real.
Electromagnetic CLFV interactions similar to \eqref{eq:Ldpl} can be obtained by replacing $A'$ with a photon. However, such dangerous CLFV interactions could 
be suppressed when electrically neutral CP-even and odd scalar propagate in loop 
as discussed in Ref.~\cite{Araki:2021vhy}.

Given the Lagrangian in Eq. (\ref{eq:Ldpl}), the total decay width of $A'$ is given as follows:
\begin{align}
\Gamma_{\mathrm{total}} = 
\sum_{\ell=e,\mu,\tau}\Gamma(A' \rightarrow \ell\bar{\ell})
+ \Gamma(A' \rightarrow e\bar{\mu})
+ \Gamma(A' \rightarrow \mu\bar{e})~.
\end{align}
The partial decay width into charged leptons is written as
\begin{align}
\Gamma(A' \rightarrow \ell\bar{\ell'}) 
&= \frac{D^2}{12 \pi} m_{A'}^3 ~\lambda\left(\frac{m_\ell^2}{m_{A'}^2}, \frac{m_{\ell'}^2}{m_{A'}^2}\right) \nonumber \\
&\quad \times 
\left[
\frac{1}{2} 
+ \frac{1}{2}\frac{m_\ell^2 + 6m_\ell m_{\ell'} + m_{\ell'}^2}{m_{A'}^2}
- \frac{(m_\ell^2 - m_{\ell'}^2)^2}{m_{A'}^4}
\right],
\end{align}
where $m_{A'}$ stands for the mass of $A'$, and $D = \mu_\ell$ or $\mu'$ for CLFC or CLFV decays, respectively.

\section{DETECTION OF CLFV DECAYs at FASER}
\label{sec:prod-detect}

From the upstream of the beam axis, the FASER detector \cite{FASER:2022hcn} consists of the neutrino detector (FASER$\nu$), the FASER scintillator veto station, the decay volume, the timing scintillator station, the FASER tracking spectrometer \cite{FASER:2021ljd}, the pre-shower scintillator system, and the electromagnetic (EM) calorimeter system. The detector also includes three 0.57~T dipole magnets, one surrounding the decay volume and the other two embedded in the tracking spectrometer.
The trigger and data acquisition system of the FASER is summarized in Ref.~\cite{FASER:2021cpr}.

To discover decays of the light bosons into $e\mu$, identification of an electron and a muon is crucial. The silicon strip detector used for the spectrometer has the position resolution of 16~$\mu$m in the precision coordinate and 816~$\mu$m in the other coordinate for a single hit. The 0.57~T magnetic field can separate $e\mu$ with opposite charges and realize the momentum resolution of $\sim$5\% for 500~GeV with nine layers of the silicon strip detector \cite{Ariga:2018pin}. For the signal candidate, it will be required that two tracks originate from the same vertex and have the same momentum.

The electromagnetic calorimeter in FASER can measure particle energy with better than 1\% resolution for 1~TeV electrons. Even though segmentation of the calorimeter is only four, the excellent energy resolution provides capability to identify the electron-muon final state. An electron deposits all energy in the calorimeter. On the other hand, a muon loses only energy of the Minimum Ionizing Particle (MIP) in 66 scintillator layers with 4~mm thickness in the calorimeter, and the total energy deposit is negligible in comparison with that of an electron. For that reason, half of deposit energy compared to the total momentum of two tracks is indication of the signal events. The pairs of electrons and muons produced from photons in radiative muon events that are one of the main backgrounds for FASER may be useful for in-situ calibration of the energy measurement with the calorimeter.

The silicon strip module used for the spectrometer has more than 99\% of detection efficiency \cite{Campabadal:2005cn}. Considering about nine silicon strip layers in the spectrometer, the detection efficiency can be assumed as 100\% for the signal events. The calorimeter also has 100\% efficiency for an electron and muon.

Rejection of the background coming from outside of the FASER detector is important to obtain clear signature of the signal event. Although natural rock and LHC shielding can eliminate most of potential backgrounds, there still remain high energetic muons with radiation and neutrinos as the main backgrounds. In the simulation, 80k muon events with $\gamma$, electro-magnetic or hadronic shower as well as a few neutrino events with charge current or neutral current interaction are expected to enter the FASER detector from the direction of the IP with energy of secondary particles above 100~GeV in 150~fb$^{-1}$ \cite{FASER:2018ceo}. Assuming 99.99\% veto efficiency of each scintillator station, these backgrounds can be reduced to negligible level. For that reason, it can be assumed that the signal is identified with almost 100\% probability by utilizing measurement of the vertex and momentum of two tracks and total energy deposit in the calorimeter.

For more precise identification of the electron-muon event, the pre-shower scintillator system will be replaced by the interleaved pixel sensors and tungsten layers in 2024 \cite{Boyd:2803084}. The pixel sensor with hexagonal pixels of 65~$\mu$m side realizes good separation capability of an individual electromagnetic shower and accordingly discrimination of an electron and muon.

Upgrade of the FASER detector (FASER2) is also planned to extend sensitivity to new particles in the operation at the HL-LHC. The FASER2 detector will be installed in the Forward Physics Facility (FPF) which will be constructed 
to situate several experiments to utilize the forward proton-proton interactions at the HL-LHC \cite{Feng:2022inv}. The detector will be enlarged to increase statistics hundreds times larger than FASER, keeping the detector performance. Table \ref{tab:faser-dimension} summarizes the places and dimensions of the FASER and FASER2 detector as well as the integrated luminosities, that are assumed in this study.

\section{Production and number of events}
\label{sec:production}
In this section, we discuss production of the light bosons introduced in section~\ref{sec:Lagrangian} and show formulae to calculate the number of CLFV events at FASER.
The production mechanisms and the formulae considered in this paper are different for the (pseudo)scalar and gauge bosons.

As explained below, we consider the production from $B$ meson decays.
In our calculations, we exploit the data sets of differential cross
sections, momenta, and angles of $B$ mesons provided in the FORESEE package \cite{Kling:2021fwx}.

\subsection{Scalar- and Pseudoscalar-type interaction}
In the case of the scalar- and the pseudoscalar-type interaction, the scalar boson $\phi_l$ and ALP $a$ are produced from meson decays through the mixing with the SM Higgs boson and the direct couplings to the SM fermions, respectively. Among the meson decays, dominant production processes are those of $B$ and $K$ mesons \cite{Feng:2017vli}.
On one hand, $B$ mesons are very short-lived and can be assumed to decay at IP. 
On the other hand, $K$ mesons can travel macroscopic distances, so that a substantial number of $K$ mesons are absorbed or deflected by the LHC infrastructure before decaying into $\phi_l$ or $a$. 
Because of this, the production from $K$ mesons is subdominant in comparison with that from $B$ mesons.
Thus, in this work, we only consider the production from $B$ mesons~\footnote{
We have checked that the sensitivity regions, which will be shown in Sec. \ref{sec:result}, remain almost the same even if the productions from $K$ mesons are included. 
} and use the branching ratio of
\begin{align}
\label{eq:br_Btophi}
    {\rm Br}(B \to X_s \phi) \simeq 5.7 \left( 1 - \frac{m_\phi^2}{m_b^2} \right)^2 \theta_{h\phi}^2~, 
\end{align}
which is given in Ref.~\cite{Feng:2017vli} in the limit of $\theta_{h\phi} \ll 1$, where $m_b$ is the b-quark mass, and $\theta_{h\phi}$ denotes the scalar mixing angle introduced in \eqref{eq:Lscl}.

The ALP production from the $B$ meson decays is induced by the effective coupling of the ALP to the bottom and strange quarks given by~\cite{Beacham:2019nyx,Batell:2009jf}
\begin{align}
\label{eq:abs_coupling}
    \mathcal{L}_{abs} = -i \frac{c_{tt}}{\Lambda} \frac{m_t^2 m_b V_{ts}^* V_{tb}}{8 \pi^2 v^2} \ln\left(\frac{\Lambda^2}{m_t^2}\right) a \bar{s}_L b_R + {\rm H.c.} ~,
\end{align}
with $V_{ij}$ being the Cabibbo-Kobayashi-Maskawa (CKM) matrix.
The branching ratio of the decay, $B \to X_s a$, is obtained from this effective coupling as \cite{Batell:2009jf,Ariga:2018uku}
\begin{align}
\label{eq:br_Btoa}
    {\rm Br}(B \to X_s a) \simeq \left[ 3.1 \left( 1 - \frac{m_a^2}{m_B^2} \right) + 3.7 \left( 1 - \frac{m_a^2}{m_B^2} \right)^3 \right] \times \frac{4 v^2 c_{tt}^2}{\Lambda^2}~,
\end{align}
with $m_B$ being the $B$ meson mass.

The number of events of $S \rightarrow e\mu$, where $S=\phi_l$ or $a$, inside the FASER detector is given by
\begin{align}
\label{eq:num-of-event_scalar}
   N_S
   = \mathcal{L} \int dp_B d\theta_B 
   \frac{d\sigma_{pp \to B}}{dp_B d\theta_B }~
   {\rm Br}(B \to X_s S)~
   {\rm Br}(S \to e \mu)~
   \mathcal{P}_{S}^{\rm det}(\bm{p}_S)~,
\end{align}
where the expected integrated luminosity is written as $\mathcal{L}$, the momentum and angle of a $B$ meson are denoted as $p_B$ and $\theta_B$, respectively, and $\mathcal{P}_{S}^{\rm det}(\bm{p}_S)$ is the probability that $S$ decays inside the detector with momentum $\bm{p}_S$.
Concrete forms of $\mathcal{P}_{S}^{\rm det}(\bm{p}_S)$ are given in appendices~\ref{apdx:decay-position} and \ref{apdx:decay-probability}.
Note that the magnitude of $\bm{p}_S$ is determined from the energy-momentum conservation once that of a $B$ meson is given.

\subsection{Vector- and Dipole-type interaction}
For the cases of the vector- and the dipole-type interaction, the gauge bosons, $Z'$ and $A'$, cannot be produced directly from meson decays since they do not interact with quarks.
However, given the fact that the gauge bosons are massive, it is natural to expect the existence of a scalar boson spontaneously breaking the gauge symmetry.
Moreover, similarly to $\phi_l$, such a scalar boson is presumed to mix with the SM Higgs boson and has interactions with the SM fermions.
Based on these considerations, for the vector- and the dipole-type interaction, we further introduce the following interaction Lagrangian
\begin{align}
\label{eq:laglangian-X}
    \mathcal{L}_{\phi_g} = 
    g_Gm_G \phi_g G_\mu G^\mu + \frac{\theta_{h\phi}}{v} \sum_{f} m_f \bar{f} \phi_g f~,
\end{align}
where $G = Z'$ or $A'$, and $\phi_g$ stands for the symmetry breaking scalar boson.
In the second term, $\theta_{h\phi}$ denotes the mixing angle between the SM Higgs boson and $\phi_g$, similarly to \eqref{eq:Lscl}.
With \eqref{eq:laglangian-X}, a pair of the gauge bosons can be produced from the decay of $\phi_g$ generated via meson decays, as shown in Ref.~\cite{Araki:2020wkq}.~\footnote{
For the case where the U(1)$_{\lmlt}$ gauge boson and dark photon are directly produced by the decays of mesons, too large CLFV coupling to the electron and muon needs in order to obtain enough number of CLFV decay signals at the FASER detector, and such a CLFV coupling is excluded by the current bounds of the CLFV muon decays.
}
The two-body decay widths of $\phi_g$ into a pair of $G$ and the lighter SM fermions are given by
\begin{align}
\label{eq:phi2GG}
    \Gamma (\phi_g \to GG) &=
    \frac{g_G^2}{8\pi} \frac{m_G^2}{m_\phi} \left( 2 + \frac{m_\phi^4}{4 m_G^4} \left( 1 - \frac{2 m_G^2}{m_\phi^2} \right)^2 \right)
    \sqrt{1 - \frac{4 m_G^2}{m_\phi^2}}~, \\
    \Gamma (\phi_g \to f\bar{f}) &=
    \frac{m_\phi}{8 \pi} \left( \frac{m_f}{v} \right)^2 \theta_{h\phi}^2 \left( 1 - \frac{4 m_f^2}{m_\phi^2} \right)
    \sqrt{1 - \frac{4 m_f^2}{m_\phi^2}}~.
\label{eq:phi2ff}
\end{align}
It should be stressed that, in \eqref{eq:phi2GG}, the decay width is enhanced due to the longitudinal mode by the factor of $m_\phi^2/m_G^2$ for $m_\phi \gg m_G$.
In analogy with the scalar-type interaction, we only consider the production from the $B$ meson decays in this work.
Note that we use the common symbols $m_\phi$ and $\theta_{h\phi}$ for both $\phi_l$ and $\phi_g$.

The number of events of $G \rightarrow e\mu$ with $G=Z', A'$ inside the FASER detector is given by
\begin{align}
\label{eq:num-of-event_gauge}
   N_G
   &= \mathcal{L} \int dp_B d\theta_B 
   \frac{d\sigma_{pp \to B}}{dp_B d\theta_B } {\rm Br}(B \to X_s \phi_g) {\rm Br}(\phi_g \to G_1 G_2)
   \sum_{j=1,2} {\rm Br}(G_j \to e \mu) \mathcal{P}_{G_j}^{\rm det}(\bm{p}_{G}, \bm{p}_\phi)~,
\end{align}
where $\mathcal{P}_{G_j}^{\rm det}(\bm{p}_{G}, \bm{p}_\phi)$ is the probability that $G$ decays inside the detector with momentum $\bm{p}_G$ and $\bm{p}_\phi$, respectively. 
Concrete forms of $\mathcal{P}_{G_j}^{\rm det}(\bm{p}_{G}, \bm{p}_\phi)$ are given in appendices~\ref{apdx:decay-position} and \ref{apdx:decay-probability}.

\section{Result}
\label{sec:result}
\begin{table}[t]
\begin{tabular}{|c|c|c|c|c|} \hline
 \hspace{2cm} & ~~~$L_{\rm min}$~(m)~~~ & ~~~$L_{\rm max}$~(m)~~~ & ~~~$R$~(m)~~~ & ~~~$\mathcal{L}$~(ab$^{-1}$)~~~ \\ \hline \hline 
 FASER  & 478.5 & 480 & 0.1 & 0.15 \\ \hline
 FASER~2 & 475 & 480 & 1.0 & 3.0 \\ \hline
\end{tabular}
\caption{
Dimensions of the FASER and the FASER2 detector and the integrated luminosity $\mathcal{L}$ used in this study. 
Here, $L_{\rm min}$ and $L_{\rm max}$ are the distances between IP and the front and the rear of the detector, respectively, and $R$ is the radius of the detector. 
}
\label{tab:faser-dimension}
\end{table}
In this section, we calculate the expected number of events of the CLFV decays at FASER based on Eqs.~(\ref{eq:num-of-event_scalar}) and (\ref{eq:num-of-event_gauge}), and then derive 95\% C.L. sensitivity regions for each interaction.
As mentioned in section~\ref{sec:prod-detect}, FASER can be assumed to be almost background free.
Hence, we regard parameter regions predicting more than three events as the 95\% C.L. sensitivity ones.
Nevertheless, just in case, the lower cut of $> 100$ GeV is placed on the momentum of the light bosons to reduce unexpected backgrounds.
We consider the setup of FASER2, since we find that $> 3$ events cannot be obtained with the setup of FASER due to smaller dimensions of the detector and integrated luminosity.
Details of the FASER and FASER2 setup are summarized in table~\ref{tab:faser-dimension}.

In order to compare FASER's sensitivity with the current experimental limits, we also show bounds from the rare muon decays: $\mu \rightarrow e\gamma$ and $\mu \rightarrow eee$, and that from the E137 electron beam dump experiment \cite{Bjorken:1988as}.
For the muon decays, we impose Br$(\mu \rightarrow e\gamma) < 4.2 \times 10^{-13}$ \cite{MEG:2016leq} and Br$(\mu \rightarrow eee) < 10^{-12}$ \cite{SINDRUM:1987nra} and derive upper bounds by using formulas given in Refs.~\cite{Lindner:2016bgg} and \cite{Araki:2021vhy}, respectively.
As for E137, we derive exclusion regions by following the analyses presented in Ref.~\cite{Araki:2021vhy}.
Note that there also exist bounds from $\mu\rightarrow eX$ \cite{Derenzo:1969za,TWIST:2014ymv,PIENU:2020loi,Calibbi:2020jvd} in mass regions below $m_\mu - m_e$.
These bounds are not shown since we are interested in CLFV decays into $e\mu$, that is, mass regions above $m_\mu + m_e$.

\subsection{Scalar-type interaction}
\begin{figure}[t]
\centering
\includegraphics[width=0.54\textwidth]{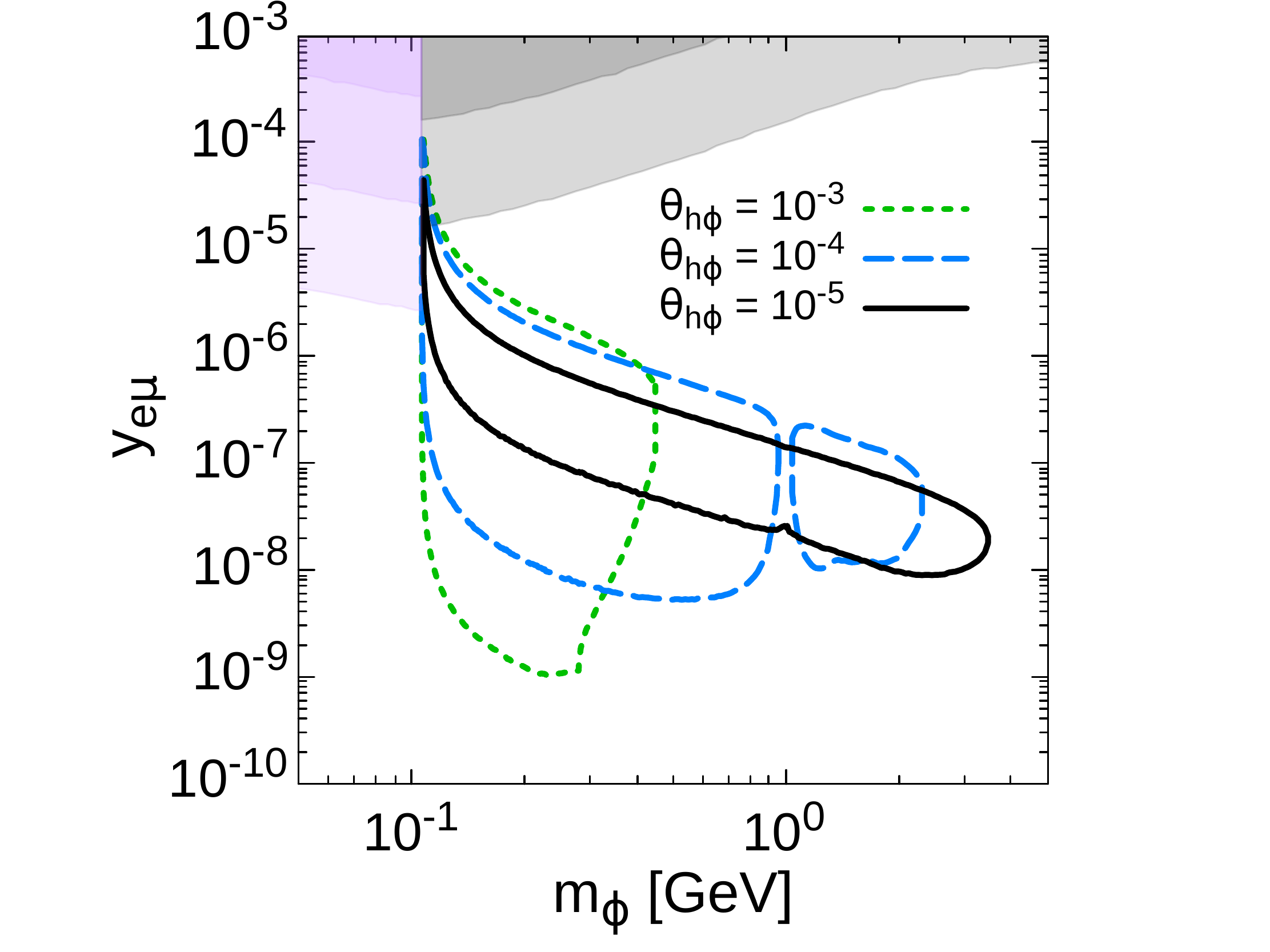}
\hspace{-16mm}
\includegraphics[width=0.54\textwidth]{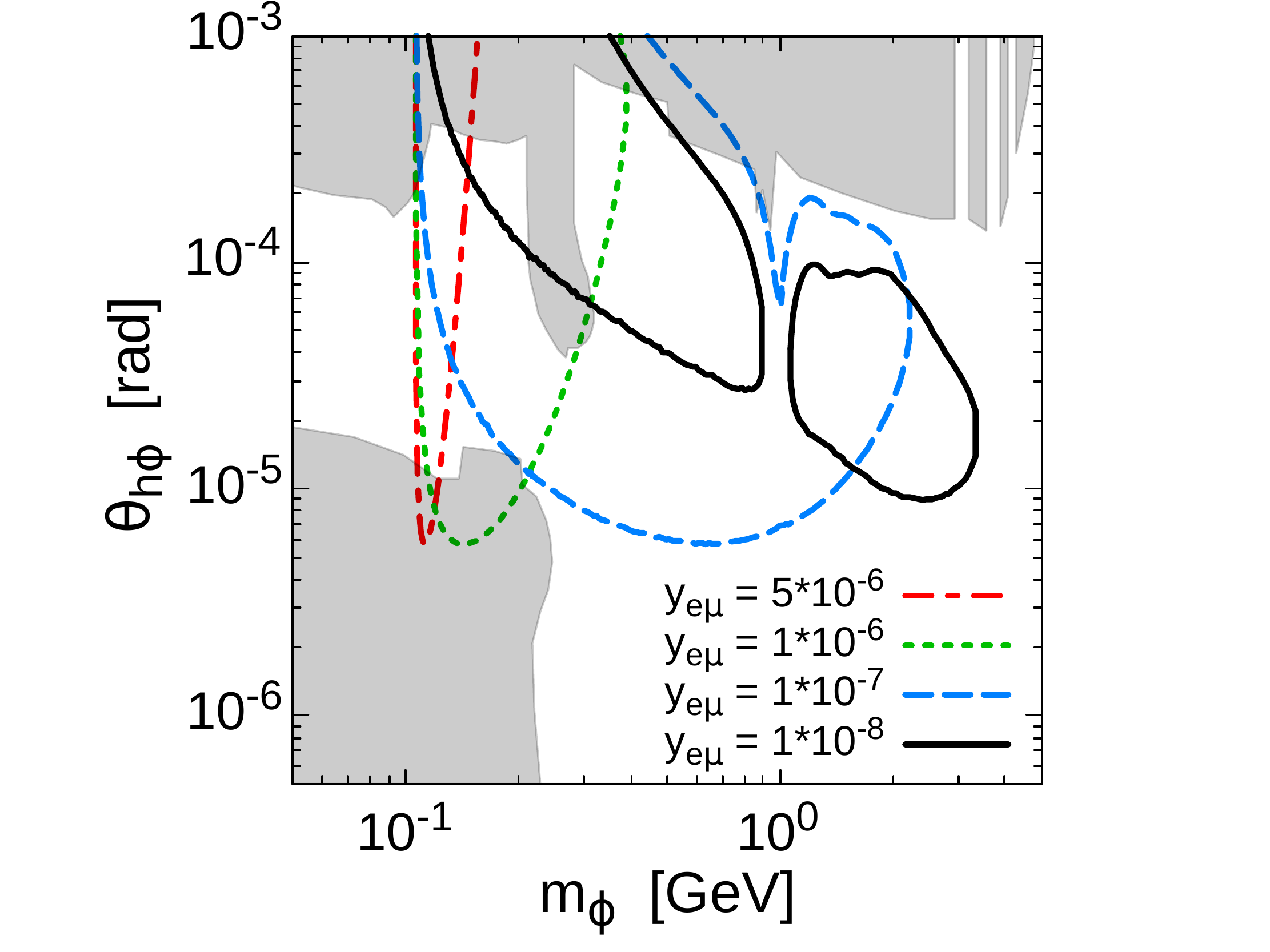}
\caption{
The contour plots of 95 \% C.L. sensitivity regions for the scalar-type interaction in the 
$m_\phi - y_{e\mu}$ plane (left) and the $m_\phi - \theta_{h\phi}$ plane (right).
Left: the purple shaded areas are excluded by E137 for $\theta_{h\phi} = 10^{-3}$ (light), $10^{-4}$ (medium), and $10^{-5}$ (dark).
The gray shaded areas are excluded by $\mu \rightarrow e\gamma$ for $\theta_{h\phi} = 10^{-3}$ (light) and $10^{-4}$ (medium).
Right: the gray shaded areas show the current exclusion regions for the dark-Higgs model.
}
\label{fig:dh95}
\end{figure}
We show numerical results for the scalar-type interaction given in \eqref{eq:Lscl}. For simplicity, we assume that $y_{e\mu}=y_{\mu e}$, and they are real.
Then, only three model parameters remain: the scalar boson mass $m_\phi$, the CLFV coupling constant $y_{e\mu}$, and the scalar mixing angle $\theta_{h\phi}$. 
In the left panel of figure \ref{fig:dh95}, we present the $95$\% C.L. sensitivity regions in the $m_\phi - y_{e\mu}$ plane for $\theta_{h \phi} = 10^{-3}$ (green dotted), $10^{-4}$ (blue dashed), and $10^{-5}$ (black solid) as illustrating examples. 
In the top-left of the figure, regions excluded by E137 are shown as the purple shaded areas for $\theta_{h\phi} = 10^{-3}$ (light), $10^{-4}$ (medium), and $10^{-5}$ (dark) \cite{Araki:2021vhy}. These exclusion regions are closed at $m_\phi = m_\mu + m_e$, at which $\phi_l \rightarrow e\mu$ opens and reduces the branching ratio of $\phi_l \rightarrow ee$.
Above this threshold, there are exclusion regions by $\mu \rightarrow e\gamma$. The exclusion regions are depicted as the gray shaded areas for $\theta_{h\phi} = 10^{-3}$ (light) and $10^{-4}$ (medium); The bound for $\theta_{h\phi} = 10^{-5}$ exists just above $y_{e\mu} = 10^{-3}$.
Note that, in the case of the scalar-type interaction, bounds from $\mu \rightarrow eee$ are weaker than those from $\mu \rightarrow e\gamma$ because Br($\mu \rightarrow eee$) is suppressed by the electron mass in $y_{ee}=(\theta_{h\phi} m_e) /v$ in comparison with Br($\mu \rightarrow e\gamma$). 
From the figure, one can see that the sensitivity region extends into the smaller coupling and lighter mass region for a larger $\theta_{h \phi}$. The reason can be understood as follows. 
For a larger $\theta_{h \phi}$, more scalar bosons are produced from $B$ meson decays because 
the decay rate \eqref{eq:br_Btophi} is proportional to $\theta_{h \phi}^2$. Such a large production increases the sensitivity in the smaller coupling region. 
On the other hand, the CLFC decay widths of the scalar boson into leptons and mesons also become larger, because those decays occur also through the scalar mixing. 
Then, the decay length of the scalar boson becomes shorter, and the CLFV branching ratio becomes smaller.
Furthermore, the decay length becomes significantly short above two pion mass threshold as shown in figure~1 of Ref.~\cite{Feng:2017vli}.
Combining these facts, the number of the scalar boson reaching and decaying in the detector is reduced. Then the sensitivity is decreased in the heavy mass region for a larger $\theta_{h\phi}$. 
As for a smaller $\theta_{h\phi}$, the sensitivity region becomes narrow since the production from $B$ mesons is reduced.
The sensitivity region disappears for $\theta_{h\phi} \lsim 10^{-5}$.

In the right panel of figure \ref{fig:dh95}, the sensitivity regions are shown in the $m_\phi - \theta_{h\phi}$ plane for $y_{e\mu} = 5\times 10^{-6}$ (red dotted-dashed), $10^{-6}$ (green dotted), $10^{-7}$ (blue dashed), 
and $10^{-8}$ (black solid). As reference, the current exclusion regions for the dark-Higgs model~\footnote{
Note that these bounds are obtained for the ordinary dark-Higgs model and that the bounds could be modified due to the inclusion of the CLFV coupling.} 
are superimposed; We show the bounds obtained in Ref. \cite{Winkler:2018qyg} as well as those from LSND \cite{Foroughi-Abari:2020gju} and MicroBooNE \cite{MicroBooNE:2021usw}.
The range of $y_{e\mu}$ is restricted by the decay length of $\phi_l$ and the branching ratio of $\phi_l \rightarrow e\mu$.
If $y_{e\mu}$ is too large, the scalar boson cannot reach the detector, while if $y_{e\mu}$ is too small, one cannot obtain $> 3$ events of $\phi_l \rightarrow e\mu$.
Note that the sensitivity regions are found in the large mass region for a smaller $y_{e\mu}$, simply due to the fact that the CLFV decay width is proportional to $y_{e\mu}^2 m_\phi$.
From the figures, we find that FASER2 can explore the CLFV coupling within $10^{-8} \lesssim y_{e\mu} \lesssim 10^{-4}$ if $\theta_{h\phi}$ is constrained as $\theta_{h\phi} \lesssim 10^{-4}$.

\subsection{Pseudoscalar-type interaction}

\begin{figure}[t]
\centering
\includegraphics[width=0.54\textwidth]{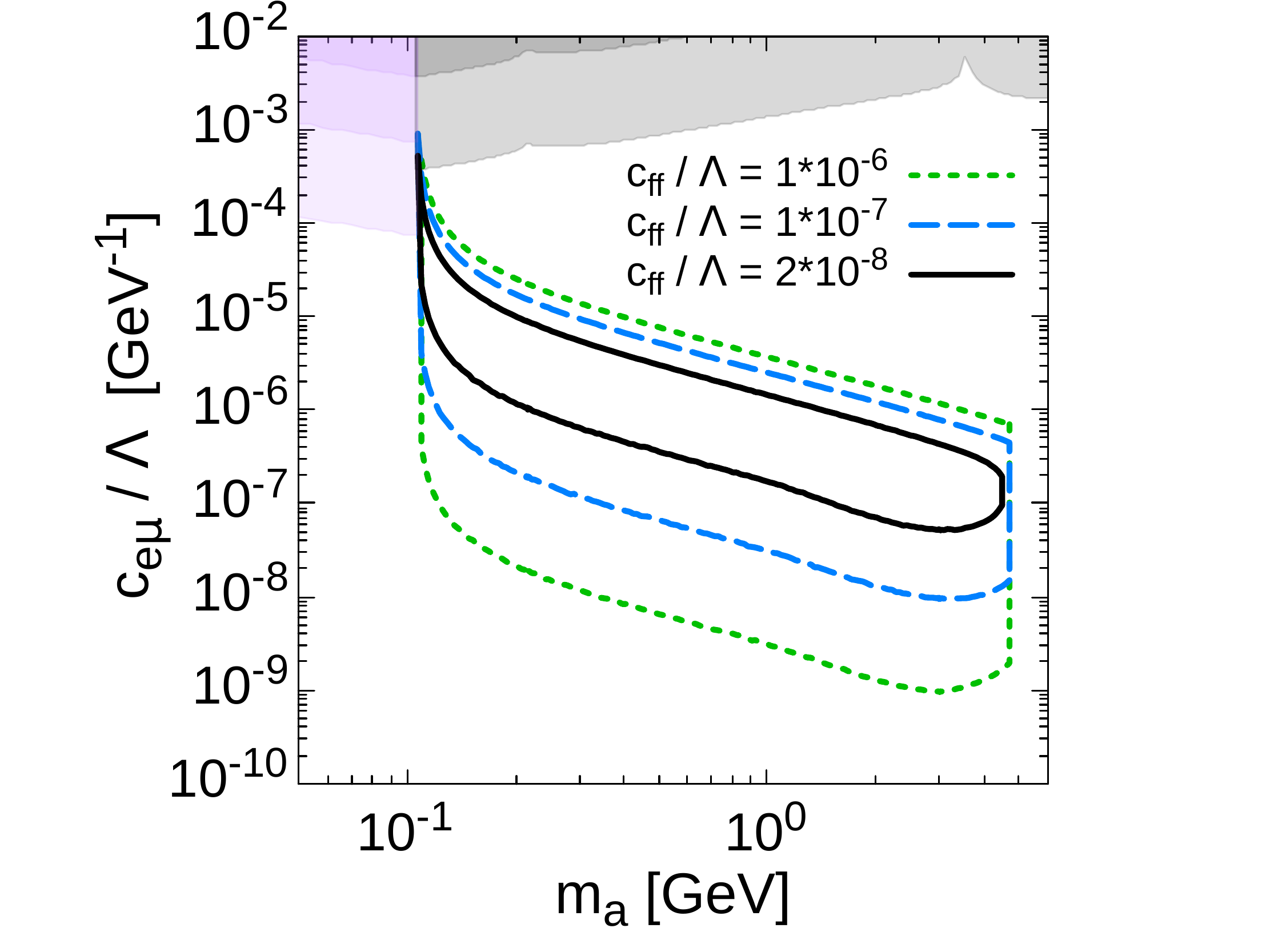}
\hspace{-16mm}
\includegraphics[width=0.54\textwidth]{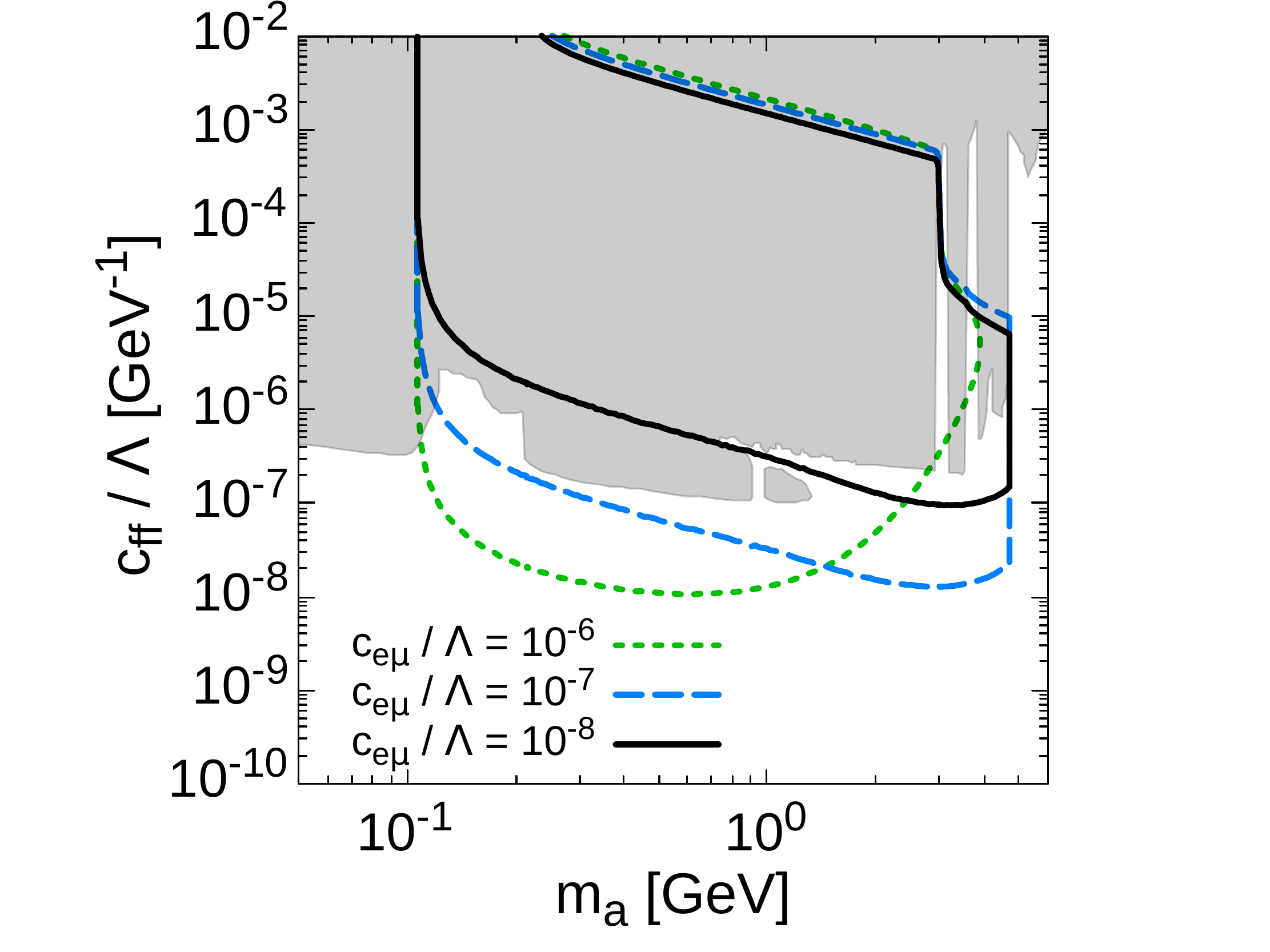}
\caption{
The contour plots of 95 \% C.L. sensitivity regions for the pseudoscalar-type interaction 
in the $m_a - c_{e\mu}/\Lambda$ plane (left) and the $m_a - c_{\ell \ell}/\Lambda$ plane (right). 
Left: the purple shaded areas are excluded by E137 for $c_{ff}/\Lambda = 10^{-6}$ (light), $10^{-7}$ (medium), and $2 \times 10^{-8}$ GeV$^{-1}$ (dark).
The gray shaded areas are excluded by $\mu \rightarrow e\gamma$ for $c_{ff}/\Lambda = 10^{-6}$ (light) and $10^{-7}$ (medium) in the case of $\Lambda = 1$ TeV.
Right: the gray shaded area shows the current exclusion region given in Ref.~\cite{Beacham:2019nyx}.
}
\label{fig:alp95}
\end{figure}
For the case of the pseudoscalar-type interaction given in \eqref{eq:Lalp}, the sensitivity regions are shown in the $m_a - c_{e\mu}/\Lambda$ plane (left panel) and $m_a - c_{ff}/\Lambda$ plane (right panel) of figure~\ref{fig:alp95}.
In the figures, for simplicity, we assume that all couplings are real and the CLFC couplings are the same for all the fermions. 
In the left panel of figure~\ref{fig:alp95}, the $95$\% C.L. sensitivity contours are shown for $c_{ff}/\Lambda = 10^{-6}$ (green dotted), $10^{-7}$ (blue dashed), and $2 \times 10^{-8}$ GeV$^{-1}$ (black solid) in the $m_a - c_{e\mu}/\Lambda$ plane.
In the right panel of figure~\ref{fig:alp95}, the contours correspond to $c_{e\mu}/\Lambda = 10^{-6}$ (green dotted), 
$10^{-7}$ (blue dashed), and $10^{-8}$  GeV$^{-1}$ (black solid) in the $m_a - c_{ff}/\Lambda$ plane.
As in the case of the scalar-type interaction, we also show the current exclusion region on $c_{ff}$~\cite{Beacham:2019nyx}. 
For the constraints, $\Lambda$ is set to $1$\,TeV in this paper.

In the left panel of figure~\ref{fig:alp95}, it is seen that the sensitivity regions extend into the smaller $c_{e\mu}/\Lambda$ region as $c_{ff}/\Lambda$ increases.
This is because the production from $B$ mesons increases, similarly to the scalar-type interaction.
In the case of ALPs, however, the sensitivity does not weaken even in the heavy mass region, because decays into two pseudoscalar mesons, such as $a\rightarrow \pi\pi$, are not allowed by $CP$ invariance.
As a result, contrary to the scalar-type interaction, the number of ALPs reaching and decaying in the detector is not suppressed even in the heavy mass region.
Note that the sensitivity regions close at around $m_a \simeq 5$ GeV, since ALPs cannot be generated by $B$ meson decays.
As shown in the right panel of figure~\ref{fig:alp95}, the sensitivity regions disappear in $m_a > 2 m_c$ for $c_{ff}/\Lambda \gtrsim 10^{-5}$, where $m_c$ is the charm quark mass.
This is because the decay process of $a \to c \bar{c}$ opens, and the lifetime of the ALP becomes too short to reach the detector.
Comparing three sensitivity regions, the larger the CLFV coupling is, the more the sensitivity region shifts to the left.
This is because the larger CLFV coupling leads shorter decay length, and the smaller ALP mass needs the ALPs to reach the detector.

\subsection{Vector-type interaction}
\begin{figure}[t]
\centering
\includegraphics[width=0.54\textwidth]{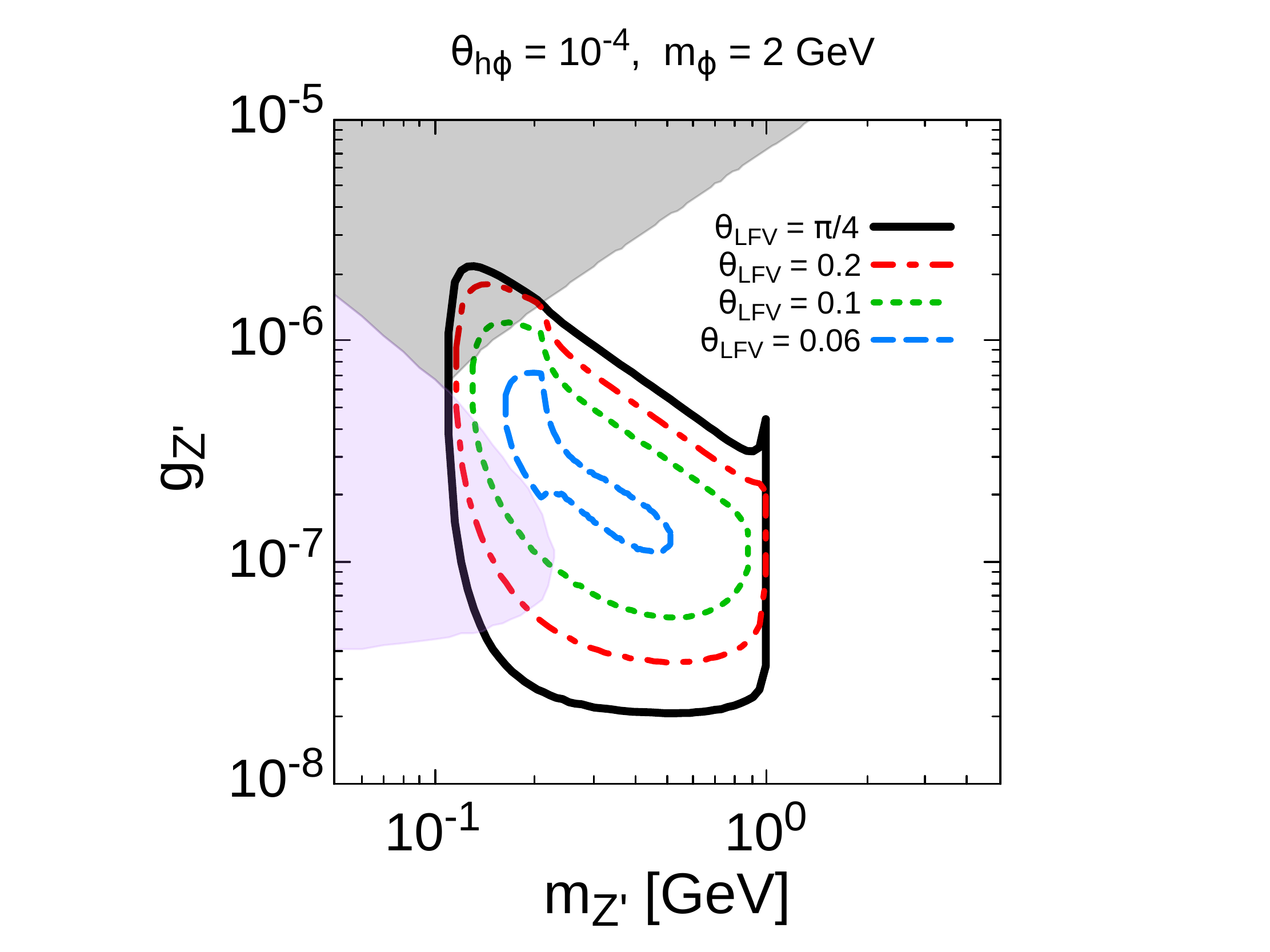}
\hspace{-16mm}
\includegraphics[width=0.54\textwidth]{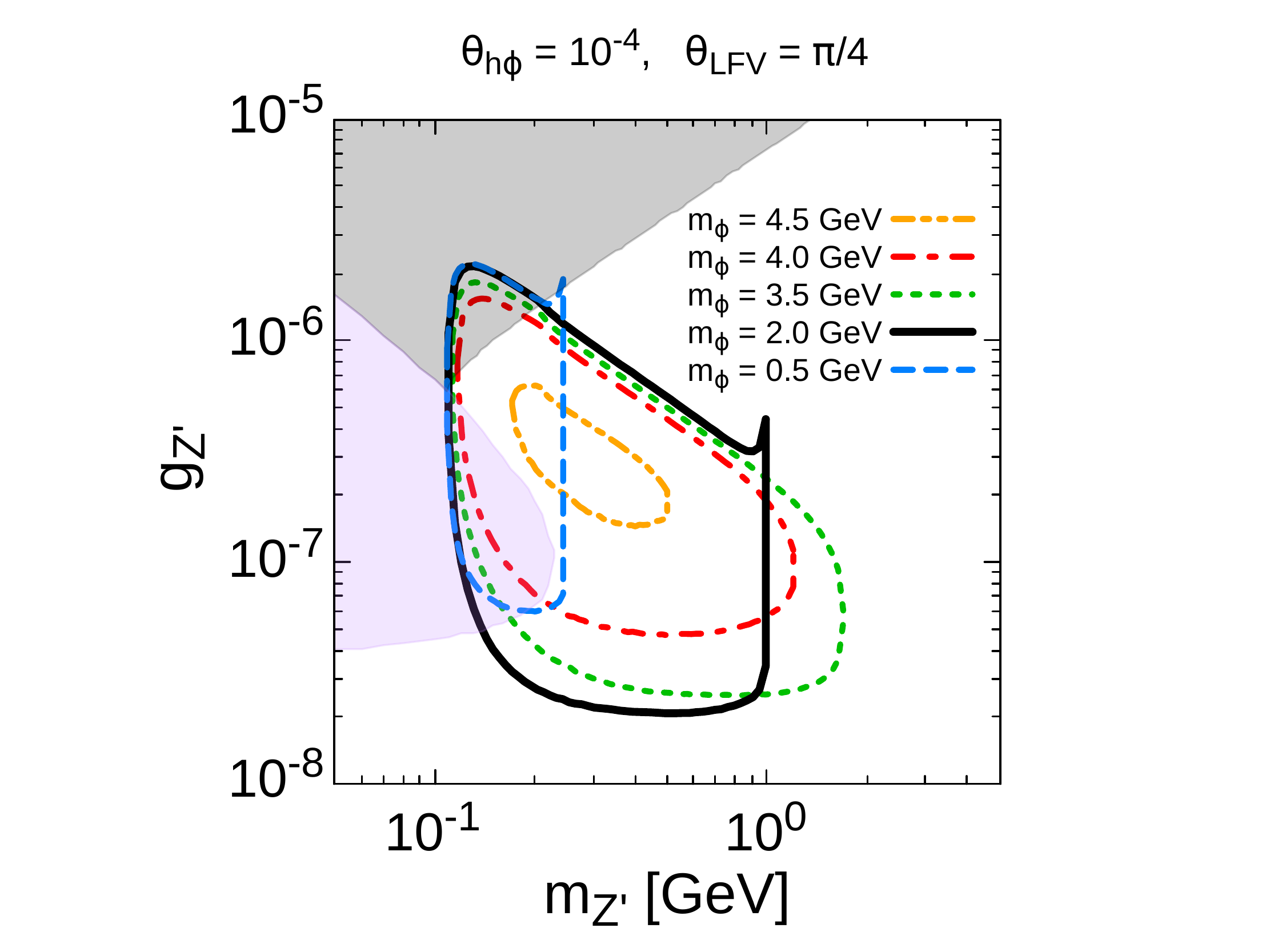}
\caption{
The contour plots of 95 \% C.L. sensitivity regions for the vector-type interaction in the $m_{Z'} - g_{Z'}$ plane.
The gray and the purple shaded area are excluded by $\mu \rightarrow eee$ and E137, respectively, for $\theta_{\rm LFV}=\pi/4$.
}
\label{fig:mt95_gzp}
\end{figure}
\begin{figure}[t]
\centering
\includegraphics[width=0.54\textwidth]{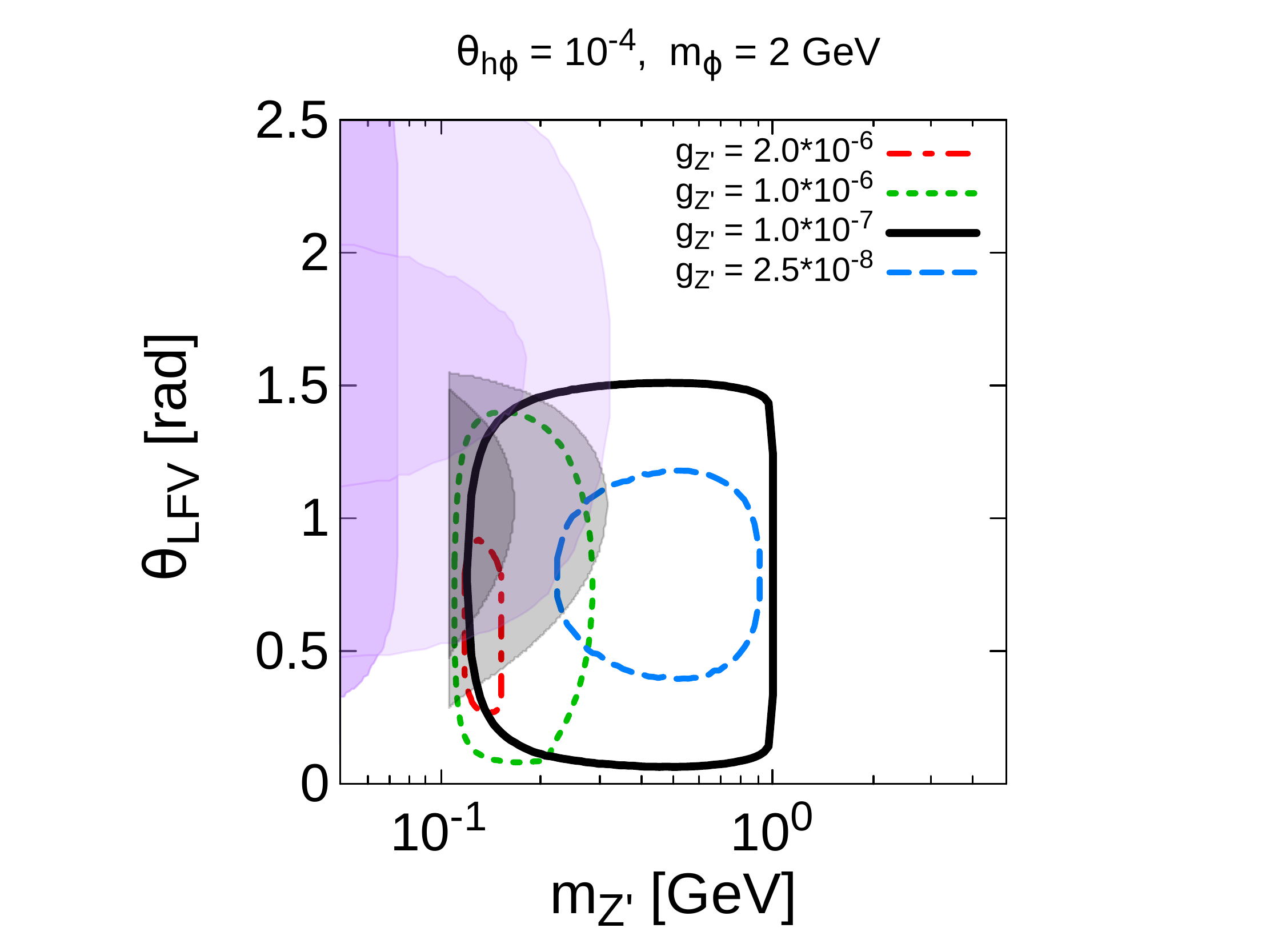}
\hspace{-16mm}
\includegraphics[width=0.54\textwidth]{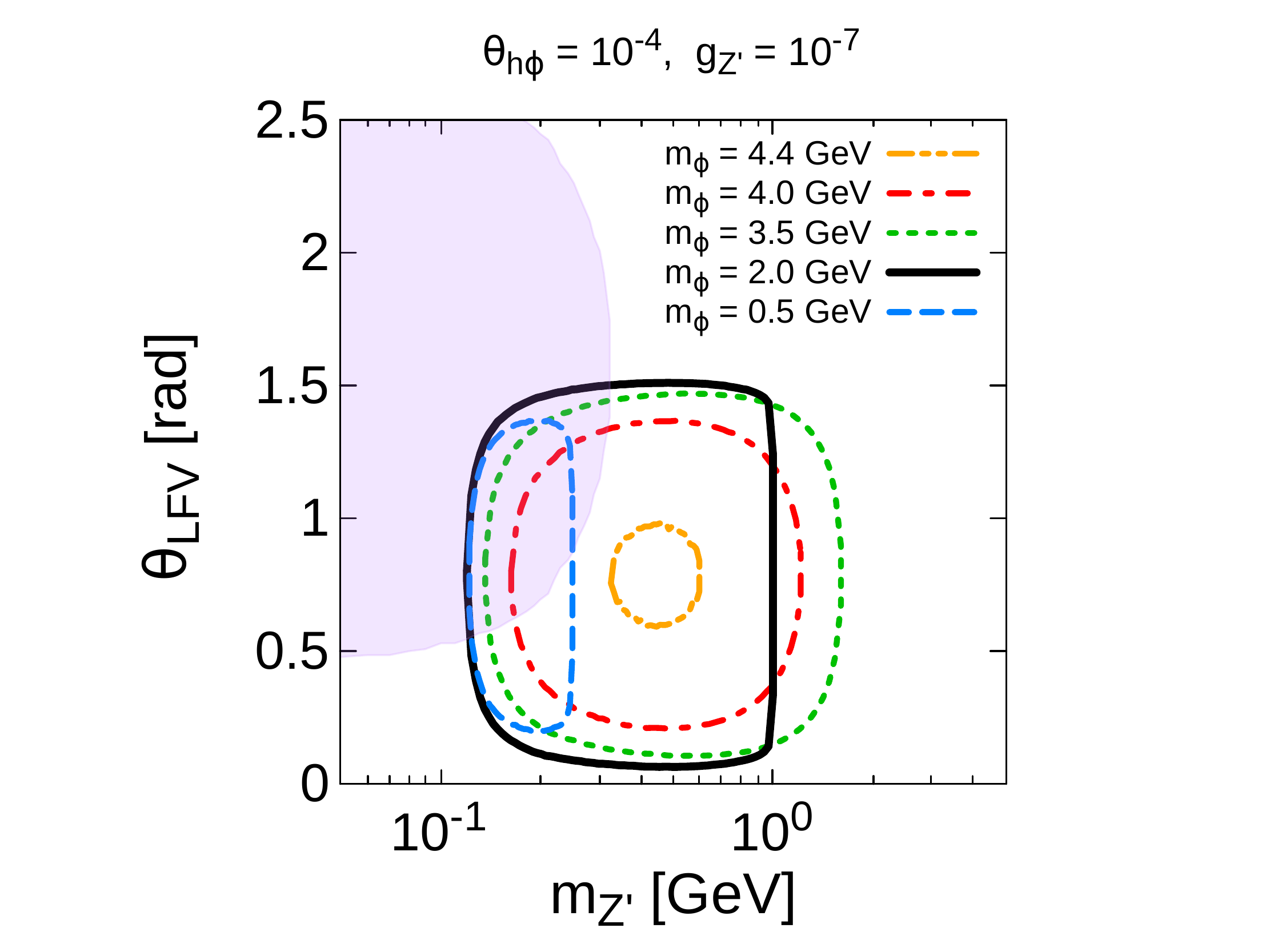}
\caption{
The contour plots of 95 \% C.L. sensitivity regions for the vector interaction in the $m_{Z'} - \theta_{\rm LFV}$ plane.
The purple shaded areas are excluded by E137 for $g_{Z'} = 10^{-7}$ (light), $2.5\times 10^{-8}$ (medium), and $10^{-6}$ (dark).
The gray shaded areas are excluded by $\mu \rightarrow eee$ for $g_{Z'}=2\times 10^{-6}$ (light) and $10^{-6}$ (medium).
}
\label{fig:mt95_mix}
\end{figure}
In the case of the vector-type interaction, the production of the gauge boson, $Z'$, is proceeded in two steps; The symmetry breaking scalar boson, $\phi_g$, is firstly produced through $B$ meson decays, and then $\phi_g$ decays into a pair of $Z'$.
There are five model parameters: $m_\phi$, $\theta_{h\phi}$, $m_{Z'}$, $g_{Z'}$, and $\theta_{\rm LFV}$.
Among these parameters, the first two parameters determine the number of $\phi_g$ produced from $B$ meson decays, while the latter three ones determine the number of signal events. 
In order to see how the sensitivity region can maximally spread, we fix $\theta_{h\phi}=10^{-4}$ in what follows.
The mass parameters are restricted to be $2 m_{Z'} < m_\phi \lesssim m_{B}$ and $m_e + m_\mu < m_{Z'} < m_\phi/2$, since $\phi_g$ ($Z'$) is required to be produced from $B$ mesons (from $\phi_g$) and decays into a pair of $Z'$ (into $e\mu$).

In order to see the sensitivity of FASER2 to the remaining parameters $g_{Z'}$ and $\theta_{\rm LFV}$, we present the 95 \% C.L. sensitivity regions in the $m_{Z'} - g_{Z'}$ plane (figure~\ref{fig:mt95_gzp}) and in the $m_{Z'} - \theta_{\rm LFV}$ plane (figure~\ref{fig:mt95_mix}).
In the left (right) panel of figure~\ref{fig:mt95_gzp}, the sensitivity regions are shown for various values of $\theta_{\rm LFV}$ ($m_\phi$) while assuming $m_\phi = 2$ GeV ($\theta_{\rm LFV}=\pi/4$).
Also, in the both panels, we show regions excluded by $\mu \rightarrow eee$ and E137 as the gray shaded and the purple shaded area, respectively, for $\theta_{\rm LFV}=\pi/4$.
The exclusion regions for the other values of $\theta_{\rm LFV}$ are much weaker than the corresponding sensitivity regions and are not shown.
In the left (right) panel of figure \ref{fig:mt95_mix}, we vary $g_{Z'}$ ($m_\phi$) while assuming $m_\phi=2$ GeV ($g_{Z'}=10^{-7}$).
In the figures, regions excluded by E137 are shown as the purple shaded areas for $g_{Z'} = 10^{-7}$ (light), $2.5\times 10^{-8}$ (medium), and $10^{-6}$ (dark); The gray shaded areas are excluded by $\mu \rightarrow eee$ for $g_{Z'}=2\times 10^{-6}$ (light) and $10^{-6}$ (medium).
Note that, in the case of the vector-type interaction, constraints from $\mu \rightarrow e\gamma$ are weaker than those from $\mu \rightarrow eee$, since Br($\mu \rightarrow e\gamma$) is suppressed by a loop factor and the electromagnetic coupling compared with Br($\mu \rightarrow eee$).

From the figures, one can see that the parameter regions of $0.05 \lesssim \theta_{\rm LFV} \lesssim 1.5$ and $10^{-8} \lesssim g_{Z'} \lesssim 10^{-6}$ can be explored at FASER2.
Here, several comments are in order.
(i) Both of the decay lengths for $\phi_g$ and $Z'$ depend on the gauge coupling $g_{Z'}$, and the range of $g_{Z'}$ is restricted to make the gauge bosons decay inside the detector.
(ii) The sensitivity region broadens as $m_\phi$ increases up to $m_\phi \simeq 3.5$ GeV, while for $m_\phi \gtrsim 3.5$ GeV the sensitivity region becomes narrower as $m_\phi$ increases.
This is because, for $m_\phi \gtrsim 3.5$ GeV, the production of $\phi_g$ from $B$ meson decays reduces as $m_\phi$ approaches the kinematical threshold of $B \rightarrow X\phi_g$.
Moreover, for $m_\phi > 2m_\tau$, $\phi_g \rightarrow \tau\tau$ opens, and the production of $Z'$ from $\phi_g$ decreases.
(iii) There are small spikes in figure \ref{fig:mt95_gzp}, e.g., around $m_{Z'} = 1$ GeV and $g_{Z'} = 4\times 10^{-7}$ in the left panel, which arise due to rapid increase of the decay length of $\phi_g$ just before closing $\phi_g \rightarrow Z'Z'$.

\subsection{Dipole-type interaction}
\begin{figure}[t]
\centering
\includegraphics[width=0.54\textwidth]{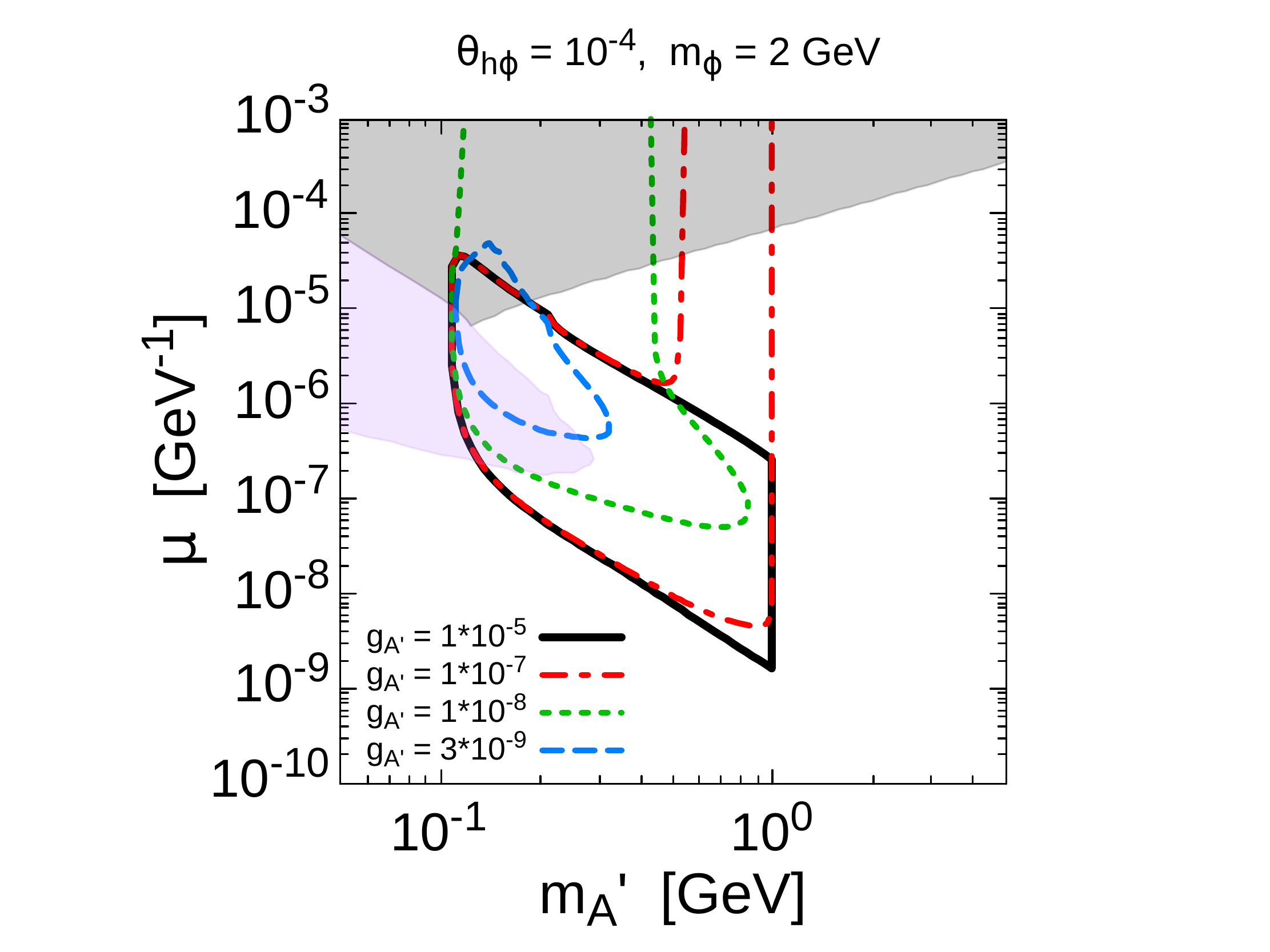}
\hspace{-16mm}
\includegraphics[width=0.54\textwidth]{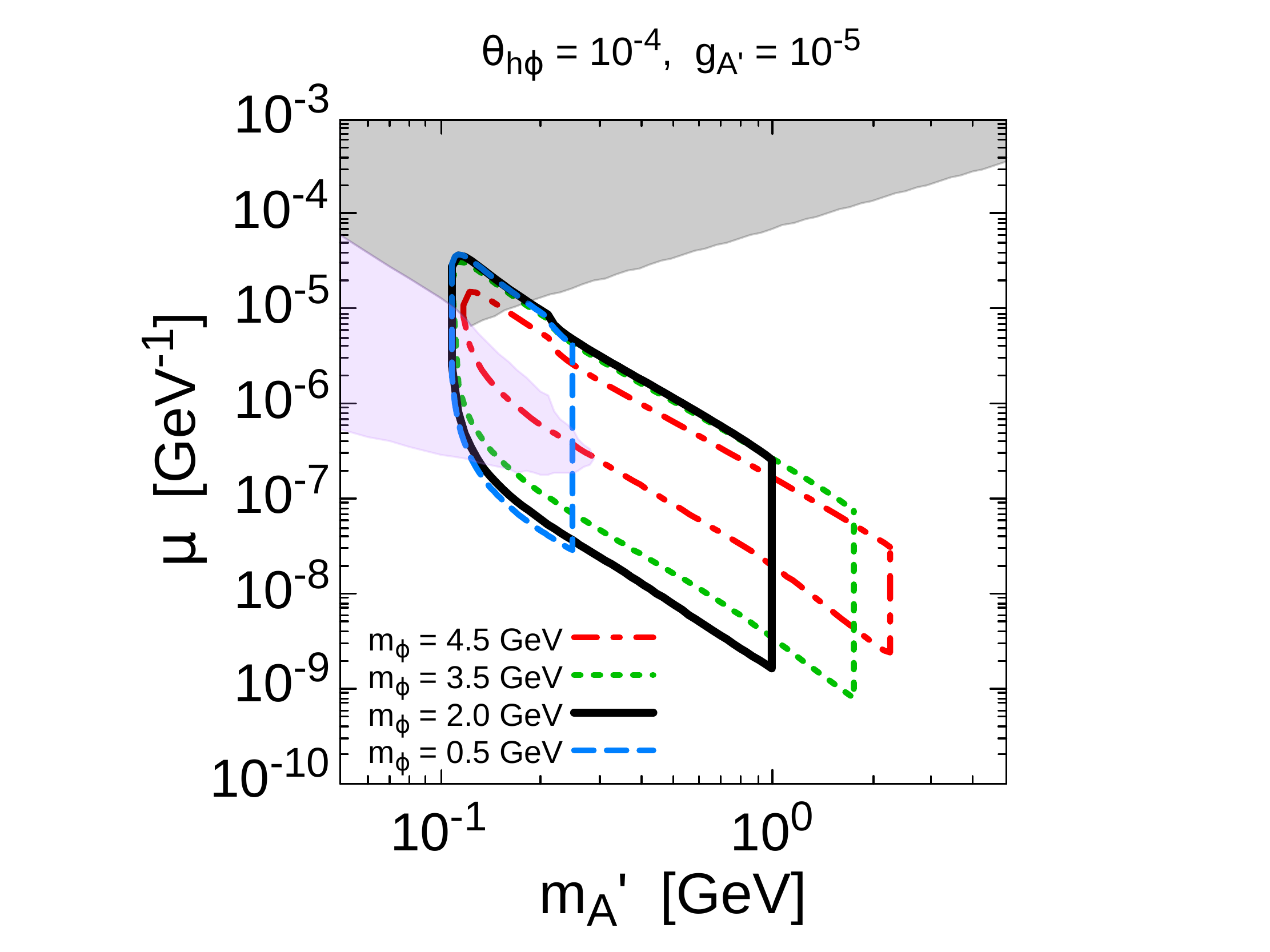}
\caption{
The contour plots of 95 \% C.L. sensitivity regions for the dipole interaction.
The gray and the purple shaded area are excluded by $\mu\rightarrow eee$ and E137, respectively.
}
\label{fig:dp95a}
\end{figure}
\begin{figure}[t]
\centering
\includegraphics[width=0.54\textwidth]{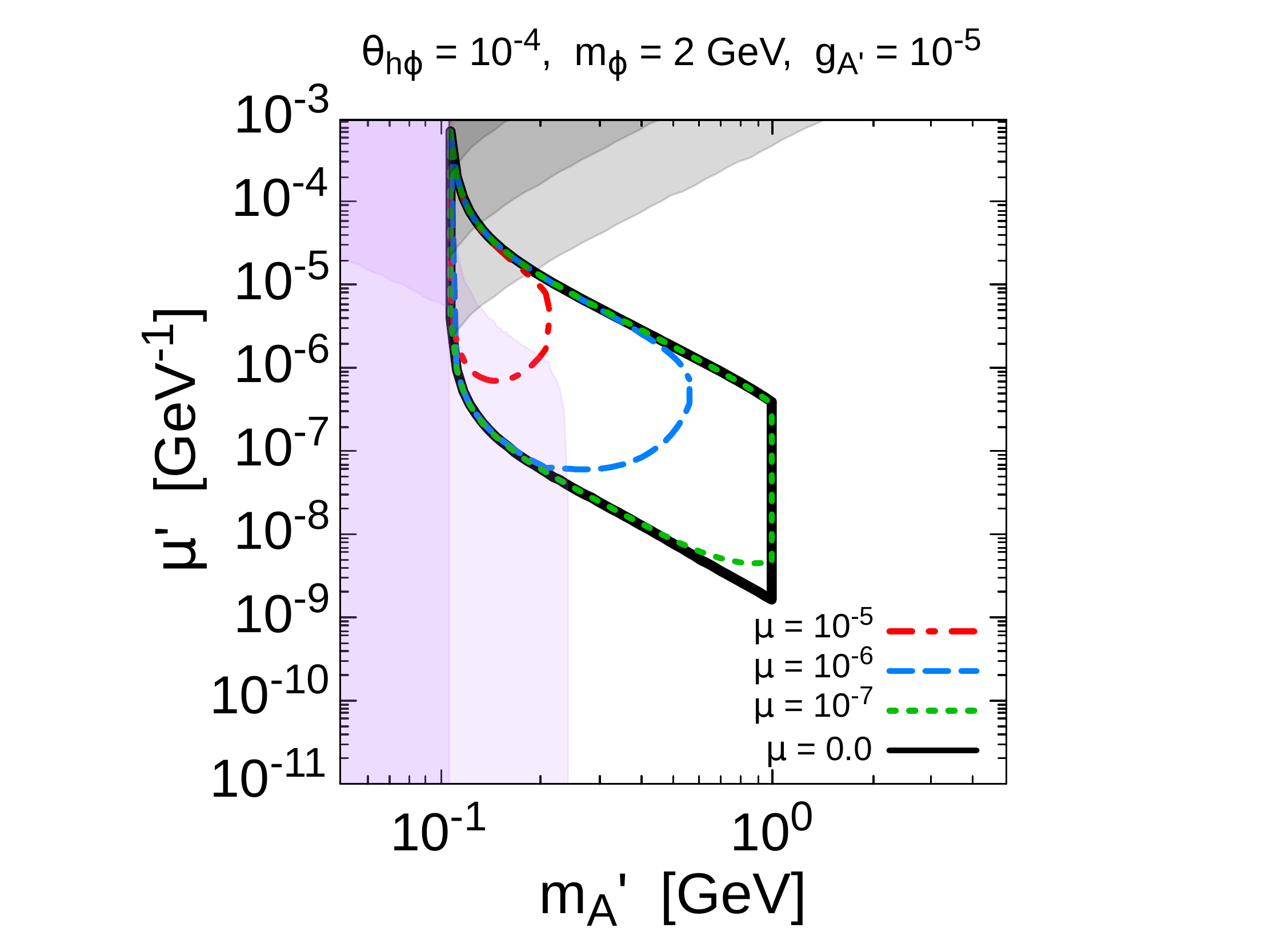}
\hspace{-16mm}
\includegraphics[width=0.54\textwidth]{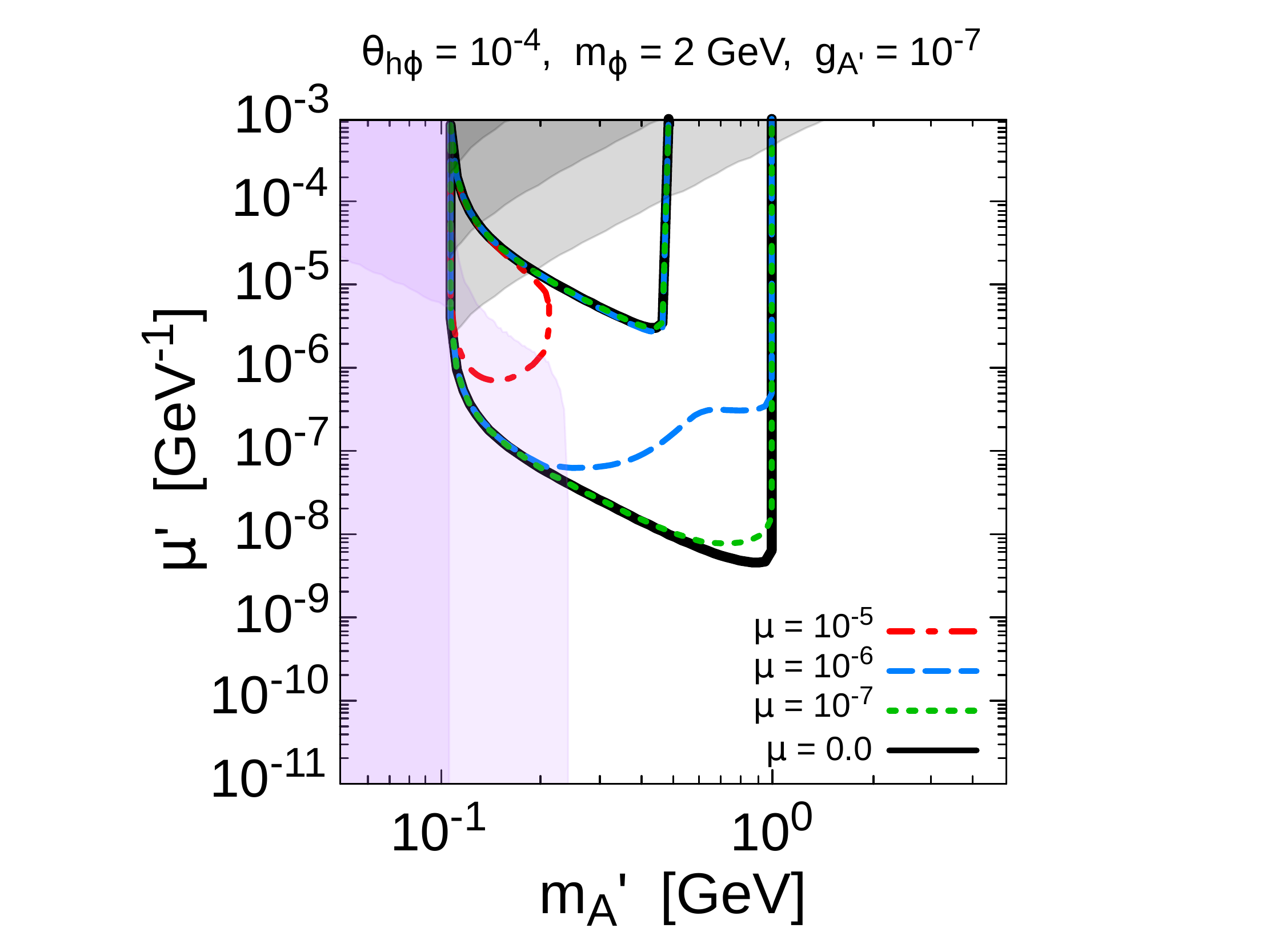}
\caption{
The contour plots of 95 \% C.L. sensitivity regions for the dipole interaction.
The gray shaded areas are excluded by $\mu\rightarrow eee$ for $\mu = 10^{-5}$ GeV$^{-1}$ (light), $10^{-6}$ GeV$^{-1}$ (medium), and $10^{-7}$ GeV$^{-1}$ (dark).
The purple shaded areas are excluded by E137 for $\mu = 10^{-5}$ GeV$^{-1}$ (medium), $10^{-6}$ GeV$^{-1}$ (light), and $10^{-7}$ GeV$^{-1}$ (dark).
}
\label{fig:dp95b}
\end{figure}
Model parameters of the dipole-type interaction are $m_\phi$, $\theta_{h\phi}$, $m_{A'}$, $g_{A'}$, $\mu'$, and $\mu_\ell$, where $\ell = e,\mu,\tau$.
As is the case for the vector-type interaction, we fix the scalar mixing as $\theta_{h\phi} = 10^{-4}$.
For simplicity, we first assume $\mu \coloneqq \mu_\ell = \mu'$.
In figure~\ref{fig:dp95a}, we show the 95 \% C.L. sensitivity regions in the $m_{A'} - \mu$ plane for various values of $g_{A'}$ (left panel) and $m_{\phi}$ (right panel).
In the figures, regions excluded by $\mu \rightarrow eee$ and E137 are also shown as the grey and the purple shaded area, respectively.
In the left panel of figure~\ref{fig:dp95a}, the sensitivity regions rise up toward the large $\mu$ region in the cases of $g_{A'} = 10^{-7}$ and $10^{-8}$.
With these values of $g_{A'}$, the symmetry breaking scalar boson, $\phi_g$, can travel macroscopic distances, and some of them can reach the detector.
As a result, the decay length of the gauge boson, $A'$, can be extremely short, and the dipole coupling $\mu$ can be very large.
Note that this behavior cannot be seen for the vector-type interaction, since a small $g_{Z'}$ results in a long decay length for not only $\phi_g$ but also $Z'$.
Below $g_{A'} \sim 10^{-8}$, the sensitivity region becomes narrow due to the decrease in the production of $A'$ via $\phi_g \rightarrow A'A'$, especially in the large $m_{A'}$ region, in which the enhancement of $m_\phi^2/m_{A'}^2$ in \eqref{eq:phi2GG} weakens.
On the other hand, for $g_{A'} \gtrsim 10^{-5}$, the sensitivity region does not change much from that of $g_{A'} = 10^{-5}$.
In the right panel of figure~\ref{fig:dp95a}, as is the case of the vector-type interaction, the sensitivity region broadens as $m_\phi$ increases up to $m_\phi \simeq 3.5$ GeV, while for $m_\phi \gtrsim 3.5$ GeV the sensitivity region becomes narrow.
From the figures, it can be found that the parameter region of $10^{-9}~{\rm GeV}^{-1} \lesssim \mu \lesssim 10^{-5}~{\rm GeV}^{-1}$ and $g_{A'} \gtrsim 10^{-9}$ can be explored by FASER2.

We next treat the CLFV coupling, $\mu'$, as an independent parameter, while keep assuming the universal CLFC interaction: $\mu \coloneqq \mu_\ell$.
In figure \ref{fig:dp95b}, we show the sensitivity regions in the $m_{A'} - \mu'$ plane for various values of $\mu$.
In the left (right) panel of figure~\ref{fig:dp95b}, we set $m_\phi=2$ GeV and $g_{A'} = 10^{-5}$ ($10^{-7}$).
In the top of the figures, the gray shaded areas display excluded regions by $\mu \rightarrow eee$ for $\mu = 10^{-5}$ GeV$^{-1}$ (light), $10^{-6}$ GeV$^{-1}$ (medium), and $10^{-7}$ GeV$^{-1}$ (dark).
Also, in the left of the figures, we represent excluded regions by E137 as the purple shaded areas for $\mu = 10^{-5}$ GeV$^{-1}$ (medium), $10^{-6}$ GeV$^{-1}$ (light), and $10^{-7}$ GeV$^{-1}$ (dark).
As can be seen from the figures, the sensitivity region becomes narrow as $\mu$ increases.
This is because both the decay length of $A'$ and BR($A' \rightarrow e\mu$) decrease as $\mu$ increases.
We find that the sensitivity region disappears for $\mu \gtrsim 10^{-5}$ GeV$^{-1}$.

\section{Summary and discussion}
\label{sec:summary}
We have explored the possibility of detecting CLFV decays at FASER, for the scalar-, pseudoscalar-, vector-, and dipole-type CLFV interaction.
The FASER detector is installed far from the collision point of ATLAS, and thus FASER has a sensitivity to long-lived particles interacting with the SM particles very weakly.
Furthermore, the FASER detector is capable of identifying an electron and a muon.
Hence, we have focused on light and weakly interacting bosons which decay into $e\mu$ in this paper.
For such a weak coupling, CLFV interactions are expected to be as large as CLFC ones although there are strong bounds on the CLFV interactions.

For the scalar- and pseudoscalar-type interactions, we have assumed that the light scalar bosons are directly produced by $B$ meson decays.
On the other hand, the U(1)$_{L_\mu-L_\tau}$ gauge bosons and the dark photons have been assumed to be produced via decays of the symmetry breaking scalar bosons.
By calculating the number of signals of the CLFV decays, we have obtained the plots of sensitivity to the CLFV interactions as shown in figures~\ref{fig:dh95}\,-\,\ref{fig:dp95b}.
We have found that, with the setup of FASER2, one can explore broad parameter regions of new physics models with CLFV couplings, which have not yet been explored by current experiments.
Comparing with other CLFV searches, FASER2 is sensitive to small CLFV coupling regions.
Therefore, FASER2 can be a complementary experiment in the exploration of CLFV processes.

The discussion in this paper can be applied to other experiments for long-lived particle searches, for example, DUNE~\cite{DUNE:2020lwj,DUNE:2020ypp}, ILC beam dump experiment~\cite{Sakaki:2020mqb,Asai:2021xtg,Asai:2021ehn,Nojiri:2022xqn}, and FACET \cite{Cerci:2021nlb}.
Moreover, searches for CLFV interactions including a tau lepton are important although the identification of a tau lepton is difficult.
We will work on these and other issues in the future.

\begin{acknowledgements}
This work was supported by JSPS KAKENHI Grant Number JP21K20365 [KA],
~JP20K04004 [YT],~JP21H00082 [YT],~JP20H01919 [HO, YT],~JP18H01210 [TA, KA, TS],~JP18K03651 [TS], and MEXT KAKENHI Grant Number JP18H05543 [KA, TS], JP22K03622 [TS].
\end{acknowledgements}

\appendix
\renewcommand{\thesubsection}{\Alph{section}.\arabic{subsection}}
\section{Decay positions}
\label{apdx:decay-position}
In this appendix, we determine an upper and a lower limit of the position integral, which will be used when calculating the decay probability in appendix~\ref{apdx:decay-probability}.

\subsection{Decay position of a gauge boson}
\begin{figure}[h]
\centering
\includegraphics[width=0.8\textwidth]{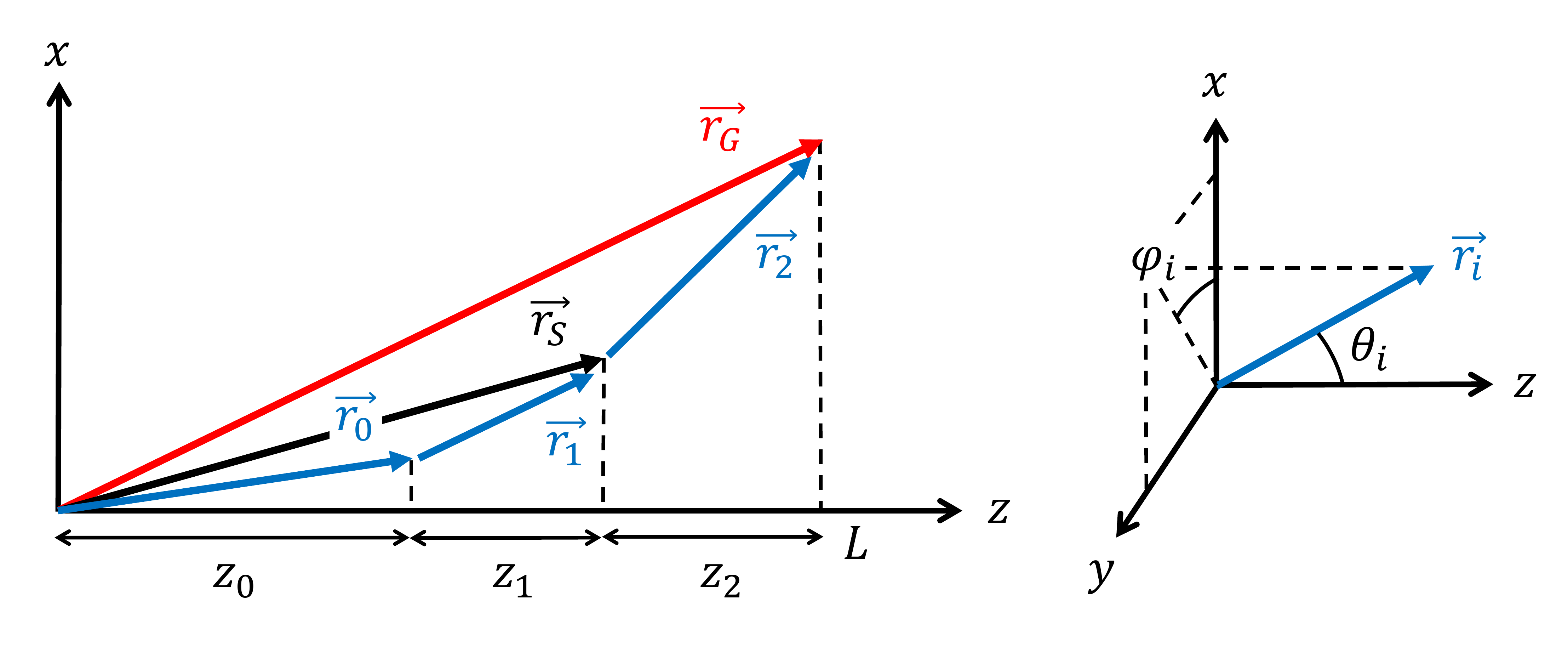}
\caption{
Left: A two-dimensional illustration of the position vectors.
Right: Definitions of the angles $\theta_i$ and $\varphi_i$ of $\Vec{r}_i$ in three dimensions.
In both the figures, the $z$ axis is taken to coincide with the LHC beam axis.
}
\label{fig:dps}
\end{figure}
We consider decay of a gauge boson $G$ in the FASER detector.
The gauge boson is assumed to be produced through the process of $X \rightarrow S \rightarrow G$, where a meson $X$ is firstly produced at IP of the ATLAS experiment and decays into a scalar boson $S$; The scalar boson, in turn, decays into the gauge boson $G$.
We define position vectors at which each particle decays as $\Vec{r}_0$, $\Vec{r}_S$, and $\Vec{r}_G$ for $X$, $S$, and $G$, respectively.
As depicted in figure \ref{fig:dps}, in terms of the position vectors, we introduce the following three vectors:
\begin{align}
& \Vec{r}_0 = z_0 \left(t_0 c_0,~t_0 s_0,~1\right)~,\\
& \Vec{r}_1 = \Vec{r}_S - \Vec{r}_0 = z_1 \left(t_1 c_1,~t_1 s_1,~1\right)~,\\
& \Vec{r}_2 = \Vec{r}_G - \Vec{r}_S = z_2 \left(t_2 c_2,~t_2 s_2,~1\right)~,
\end{align}
where $z_0$, $z_1$, and $z_2$ are the $z$ components of $\Vec{r}_0$, $\Vec{r}_1$, and $\Vec{r}_2$, respectively, and $z_2 = L - z_0 - z_1$ with $L$ being the distance to the FASER detector from IP.
Also, we have defined two angles, $\theta_i$ and $\varphi_i$, for each $\Vec{r}_i$ and used the abbreviations of $t_i = \tan\theta_i$, $c_i = \cos\varphi_i$, and $s_i = \sin\varphi_i$.
Then, the position vector of $G$ is obtained as
\begin{align}
& \Vec{r}_G = \sum_{i=0}^{2} \Vec{r}_i = (x_G,~y_G,~z_G)~,\\
& x_G = z_0 t_0 c_0 + z_1 t_1 c_1 + (L-z_1-z_0)t_2 c_2~,\\
& y_G = z_0 t_0 s_0 + z_1 t_1 s_1 + (L-z_1-z_0)t_2 s_2~,\\
& z_G = L~.
\end{align}
For $\theta_0$, we exploit the data sets provided in the FORESEE package \cite{Kling:2021fwx}, and $\varphi_0$ is set to $0$.
The other angles, $\theta_1$ and $\varphi_1$ ($\theta_2$ and $\varphi_2$), are computed by running Monte Carlo simulations in the rest frame of $X$ ($S$).
In order to make the gauge boson decay inside the detector, we impose the following condition:
\begin{align}
x_G^2 + y_G^2 - R^2 = A z_1^2 -2 B(L) z_1 + C(L) \leq 0~,
\label{eq:condition}
\end{align}
where $R$ is the radius of the FASER detector and
\begin{align}
& A = t_1^2 + t_2^2 -2 t_1 t_2 \cos(\varphi_2 - \varphi_1)~,\\
& B(L) = L \left[ t_2^2 - t_1 t_2 \cos(\varphi_2 - \varphi_1) \right]
\nonumber \\
&\hspace{10mm}
- z_0 \left[ 
  t_2^2
- t_1 t_2 \cos(\varphi_2 - \varphi_1)
- t_0 t_2 \cos(\varphi_2 - \varphi_0)
+ t_0 t_1 \cos(\varphi_1 - \varphi_0)
      \right]~,\\
& C(L) = L^2 t_2^2 - R^2
\nonumber \\
&\hspace{10mm}
+ z_0^2 \left[ t_0^2 + t_2^2 - 2 t_0 t_2 \cos(\varphi_2 - \varphi_0) \right]
- 2 z_0 L \left[ t_2^2 - t_0 t_2 \cos(\varphi_2 - \varphi_0) \right]~.
\end{align}
We require this condition at both the front and the rear of the detector, that is, at $L=L_{\rm min}$ and at $L=L_{\rm max}$, respectively.
Note that $R$, $L_{\rm min}$, and $L_{\rm max}$ are summarized in table~\ref{tab:faser-dimension}.
Given the conditions, the decay position of $S$ is restricted as $z_1^- \leq z_1 \leq z_1^+$, where $z_1^-$ and $z_1^+$ are
\begin{align}
\label{eq:z1pm}
z_1^- = 
\left\{\begin{array}{l} 
z_{1,{\rm min}}^- ~~
(~ {\rm for}~~z_{1,{\rm min}}^- > z_{1,{\rm max}}^- ~) \\ 
z_{1,{\rm max}}^- ~~
(~ {\rm for}~~z_{1,{\rm min}}^- < z_{1,{\rm max}}^- ~) 
\end{array}\right.~,~~~~
z_1^+ = 
\left\{\begin{array}{l} 
z_{1,{\rm min}}^+ ~~
(~ {\rm for}~~z_{1,{\rm min}}^+ < z_{1,{\rm max}}^+ ~) \\ 
z_{1,{\rm max}}^+ ~~
(~ {\rm for}~~z_{1,{\rm min}}^+ > z_{1,{\rm max}}^+ ~) 
\end{array}\right.~,
\end{align}
and
\begin{align}
\label{eq:def-z1pm}
& z_{1,{\rm min(max)}}^\pm 
= \frac{B\left(L_{\rm min(max)}\right) \pm 
\sqrt{B\left(L_{\rm min(max)}\right)^2-AC\left(L_{\rm min(max)}\right)}}{A}~.
\end{align}
The decay probability, which will be given in appendix~\ref{apdx:decay-probability}, is evaluated by integrating out the position integral of $S$ within this range.

\subsection{Decay position of a scalar boson}
The above discussion can easily be applied to the case of the scalar boson decays inside the detector, by setting $r_0 = 0$ and regarding $\Vec{r}_S$ and $\Vec{r}_G$ as the position vector of $X$ and $S$, respectively.
In this case, the condition in \eqref{eq:condition} constrains the decay position of $X$.
The factors of $A, B$, and $C$ in \eqref{eq:condition} are reduced to be
\begin{align}
& A = t_1^2 + t_2^2 -2 t_1 t_2 \cos(\varphi_2 - \varphi_1)~,\\
& B = t_2^2 - t_1 t_2 \cos(\varphi_2 - \varphi_1)~,\\
& C(L) = t_2^2 - \frac{R^2}{L^2}~,
\end{align}
and \eqref{eq:def-z1pm} becomes
\begin{align}
\label{eq:def-z1pm2}
& z_{1,{\rm min(max)}}^\pm 
= \frac{L_{\rm min(max)}\left[ B \pm \sqrt{B^2-AC\left(L_{\rm min(max)}\right)} \right]}{A}~.
\end{align}

\section{Decay probability}
\label{apdx:decay-probability}
Having determined the integration interval of $z_1$, we next show a probability that the gauge boson or the scalar boson decays inside the detector.

\subsection{Decay probability of a gauge boson}
\begin{table}[t]
\centering
\begin{tabular}{|c|c|c|c|}\hline
~$L_{\rm TAN}$(m)~ & ~$L_{\rm TAS}$(m)~ & ~$R_{\rm TAS}$(m)~ & ~$R_{\rm BP}$(m)~ \\\hline\hline
140 & 20 & 0.017 & 0.05 \\\hline
\end{tabular}
\caption{
Dimensions of the LHC infrastructure: $L_{\rm TAN}$ and $L_{\rm TAS}$ are the distances to the TAN neutral particle absorber and the TAS quadrupole absorber, respectively, from IP, and $R_{\rm TAS}$ and $R_{\rm BP}$ are the radiuses of TAS and the LHC beam pipe.
}
\label{tab:LHC}
\end{table}
\begin{table}[t]
\centering
\begin{tabular}{|c||c|c|c|}\hline
 & ~$\theta_0 < \theta_{\rm TAS}$~ & ~$\theta_{\rm TAS} < \theta_0 < \theta_{\rm BP}$~ & ~$\theta_{\rm BP} < \theta_0 < \pi/2$~ \\ \hline\hline
$z_{\rm abp}$ & ~$L_{\rm TAN}$ or $L_{\rm TAS}$~ & $L_{\rm TAS}$ & $R_{\rm BP}/\tan \theta_0$ \\\hline
\end{tabular}
\caption{
A summary of the integration interval of $z_0$.
The angles $\theta_{\rm TAS}$ and $\theta_{\rm BP}$ are defined as $\theta_{\rm TAS}=\arctan\left(R_{\rm TAS}/L_{\rm TAS}\right)$ and $\theta_{\rm BP}=\arctan\left(R_{\rm BP}/L_{\rm TAS}\right)$, respectively.
In the case of $\theta_0 < \theta_{\rm TAS}$, we choose $z_{\rm abp} = L_{\rm TAN}$ for neutral mesons, while $z_{\rm abp} = L_{\rm TAS}$ for charged mesons.
}
\label{tab:z0-range}
\end{table}
We define a probability that the gauge boson $G$ decays inside the detector as
\begin{align}
\mathcal{P}_G^\textrm{det}
= P_X \times \left[ P_{\rm out} + P_{\rm in} \right]~.
\label{eq:prob}
\end{align}
The probability $P_X$ is the decay probability of the meson $X$ and given by
\begin{align}
P_X = \frac{1}{d_X \cos\theta_0}\int_0^{z_{\rm abp}} dz_0 ~e^{-\frac{z_0}{d_X \cos\theta_0}}~.
\end{align}
The decay length of $X$ is denoted as $d_X$ and given by
\begin{align}
d_X = \frac{c \hbar}{\Gamma_X} \beta_X \gamma_X~,
\label{eq:dbar}
\end{align}
where $c$ is the speed of light in vacuum, $\hbar$ is the reduced Planck constant, $\Gamma_X$ stands for the total decay width of $X$, and $\beta_X \gamma_X = p_X/M_X$ is the lorentz factor of $X$ in terms of its momentum $p_X$ and mass $M_X$.
Here, mesons having long lifetime are possibly absorbed or deflected by the LHC infrastructure before decaying.
Depending on a value of $\theta_0$, we assume that mesons are absorbed by the TAN neutral particle absorber, the TAS front quadrupole absorber, or the LHC beam pipe.
Moreover, for charged mesons, we further assume that they are deflected by the superconducting quadrupole magnets located right by TAS.
The upper limit of the $z_0$ integral, that is $z_{\rm abp}$, indicates a position at which mesons are absorbed or deflected, and it is summarized in tables~\ref{tab:LHC} and \ref{tab:z0-range}.

\begin{figure}[h]
\centering
\includegraphics[width=0.8\textwidth]{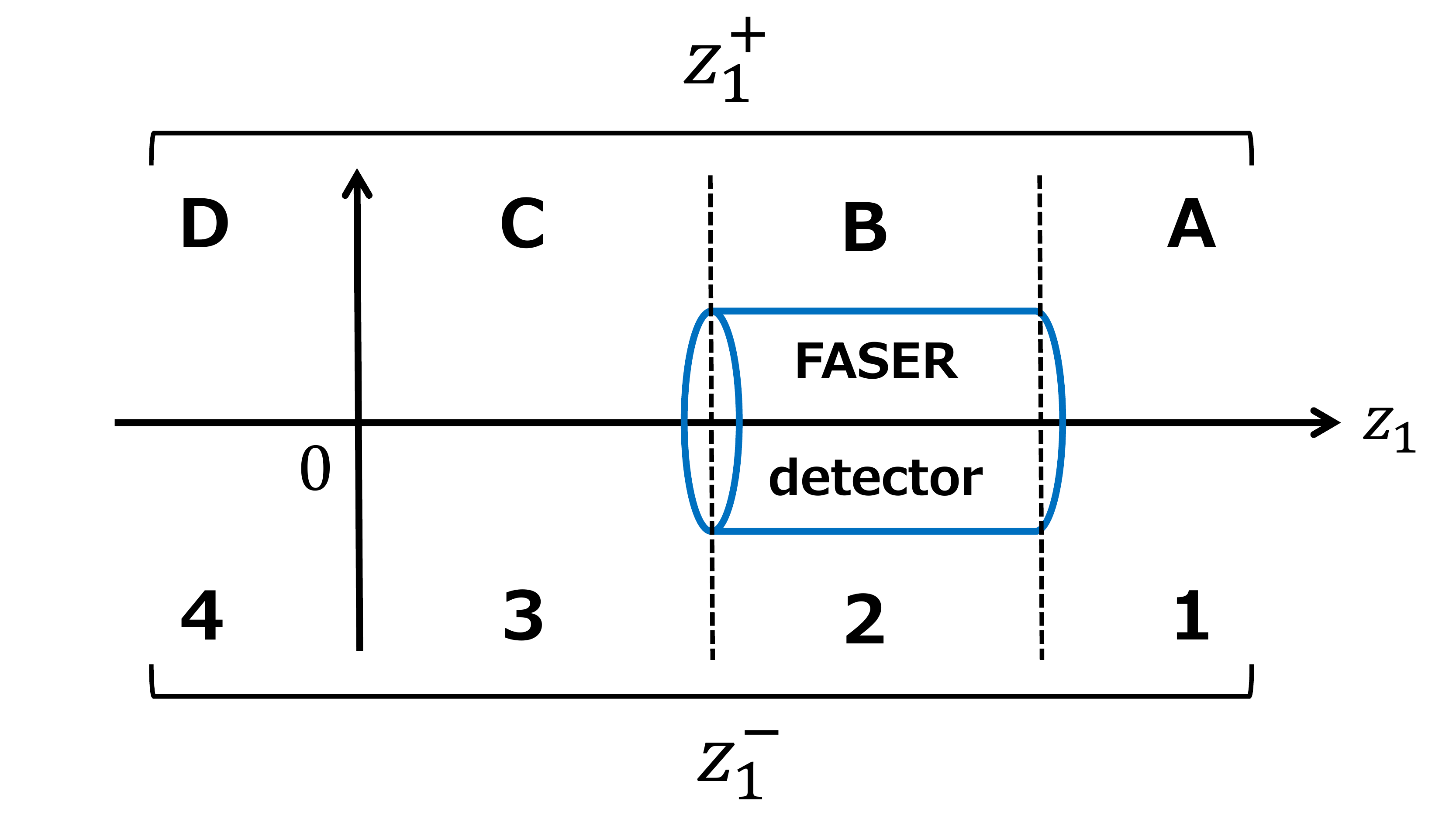}
\caption{A schematic view of the $z_1$ coordinate for the case of the gauge boson decay.
}
\label{fig:z1-range}
\end{figure}

The probabilities $P_{\rm out}$ and $P_{\rm in}$ correspond to the cases where $S$ decays outside and inside the detector, respectively, and they are products of the decay probability of $G$ and that of the scalar boson $S$:
\begin{align}
& P_{\rm out} =
\frac{1}{d_S \cos\theta_1}\int_{z_{\rm out}^{\rm lw}}^{z_{\rm out}^{\rm up}} dz_1 ~e^{-\frac{z_1}{d_S \cos\theta_1}}
\times 
\frac{1}{d_G \cos\theta_2}\int_{L_{\rm min}-z_1-z_0}^{L_{\rm max}-z_1-z_0} dz_2 ~e^{-\frac{z_2}{d_G \cos\theta_2}}~,
\label{eq:prob_out}
\\
& P_{\rm in} =
\frac{1}{d_S \cos\theta_1}\int_{z_{\rm in}^{\rm lw}}^{z_{\rm in}^{\rm up}} dz_1 ~e^{-\frac{z_1}{d_S \cos\theta_1}}
\times 
\frac{1}{d_G \cos\theta_2}\int_{0}^{L_{\rm max}-z_1-z_0} dz_2 ~e^{-\frac{z_2}{d_G \cos\theta_2}}~.
\label{eq:prob_in}
\end{align}
Definitions of the decay lengths $d_S$ and $d_G$ are similar to \eqref{eq:dbar}.
The interval of the $z_1$ integral is determined by $z_1^\pm$ obtained in Eqs.~(\ref{eq:z1pm}) and (\ref{eq:def-z1pm}).
Depending on values of $z_1^\pm$, the integral interval divides into sixteen cases.
As depicted in figure \ref{fig:z1-range}, we divide the $z_1$ coordinate into four ranges: $L_{\rm max}-z_0 < z_1$, $L_{\rm min}-z_0 < z_1 < L_{\rm max}-z_0$, $0 < z_1 < L_{\rm min}-z_0$, and $z_1 < 0$, and refer to each range as A(1), B(2), C(3), and D(4), respectively, for $z_1^+ (z_1^-)$.
Then, for instance, we label the case of $L_{\rm max}-z_0 < z_1^+$ and $L_{\rm min}-z_0 < z_1^- < L_{\rm max}-z_0$ as A-2.
Among all the possibilities, eight cases result in zero probabilities.
For the remaining eight cases, in table~\ref{tab:z1-range}, we summarize the upper and the lower limit of the $z_1$ integral.
\begin{table}[h]
\centering
\begin{tabular}{|c||c|c|c|c|c|c|c|c|}\hline
\hspace{10mm} & ~A-2~ & ~A-3~ & ~A-4~ & ~B-2~ & ~B-3~ & ~B-4~ & ~C-3~ & ~C-4~ \\\hline\hline
$z_{\rm out}^{\rm up}$ & $-$         & $L_{\rm min}-z_0$ & $L_{\rm min}-z_0$ & $-$     & $L_{\rm min}-z_0$ & $L_{\rm min}-z_0$ & $z_1^+$ & $z_1^+$ \\\hline
$z_{\rm out}^{\rm lw}$ & $-$         & $z_1^-$     & $0$         & $-$     & $z_1^-$     & $0$         & $z_1^-$ & $0$     \\\hline
$z_{\rm in}^{\rm up}$  & $L_{\rm max}-z_0$ & $L_{\rm max}-z_0$ & $L_{\rm max}-z_0$ & $z_1^+$ & $z_1^+$     & $z_1^+$     & $-$     & $-$     \\\hline
$z_{\rm in}^{\rm lw}$  & $z_1^-$     & $L_{\rm min}-z_0$ & $L_{\rm min}-z_0$ & $z_1^-$ & $L_{\rm min}-z_0$ & $L_{\rm min}-z_0$ & $-$     & $-$     \\\hline
\end{tabular}
\caption{
A summary of the integration interval of $z_1$, where $z_1^\pm$ are defined in Eqs. (\ref{eq:z1pm}).
The probabilities are zero for the other cases.
}
\label{tab:z1-range}
\end{table}

In our numerical calculations, we only consider a production from $B$ mesons.
Decay lengths of $B$ mesons are typically very short.
For example, it is estimated as
\begin{align}
d_B 
= \frac{c \hbar}{\Gamma_B}\frac{p_B}{M_B}
= c \tau_B \frac{p_B}{M_B}
\simeq 8.61\times 10^{-2}~~{\rm m}
\end{align}
for $p_B= 1$ TeV, $M_B = 5.28$ GeV, and $\tau_B = \hbar/\Gamma_B = 1.52\times 10^{-12}$ s.
Since this value is much smaller than the length between IP and the detector, we assume that $B$ mesons decay at IP and set $P_X = 1$ and $z_0 = 0$ in our numerical calculations.

\subsection{Decay probability of a scalar boson}
\begin{figure}[h]
\centering
\includegraphics[width=0.6\textwidth]{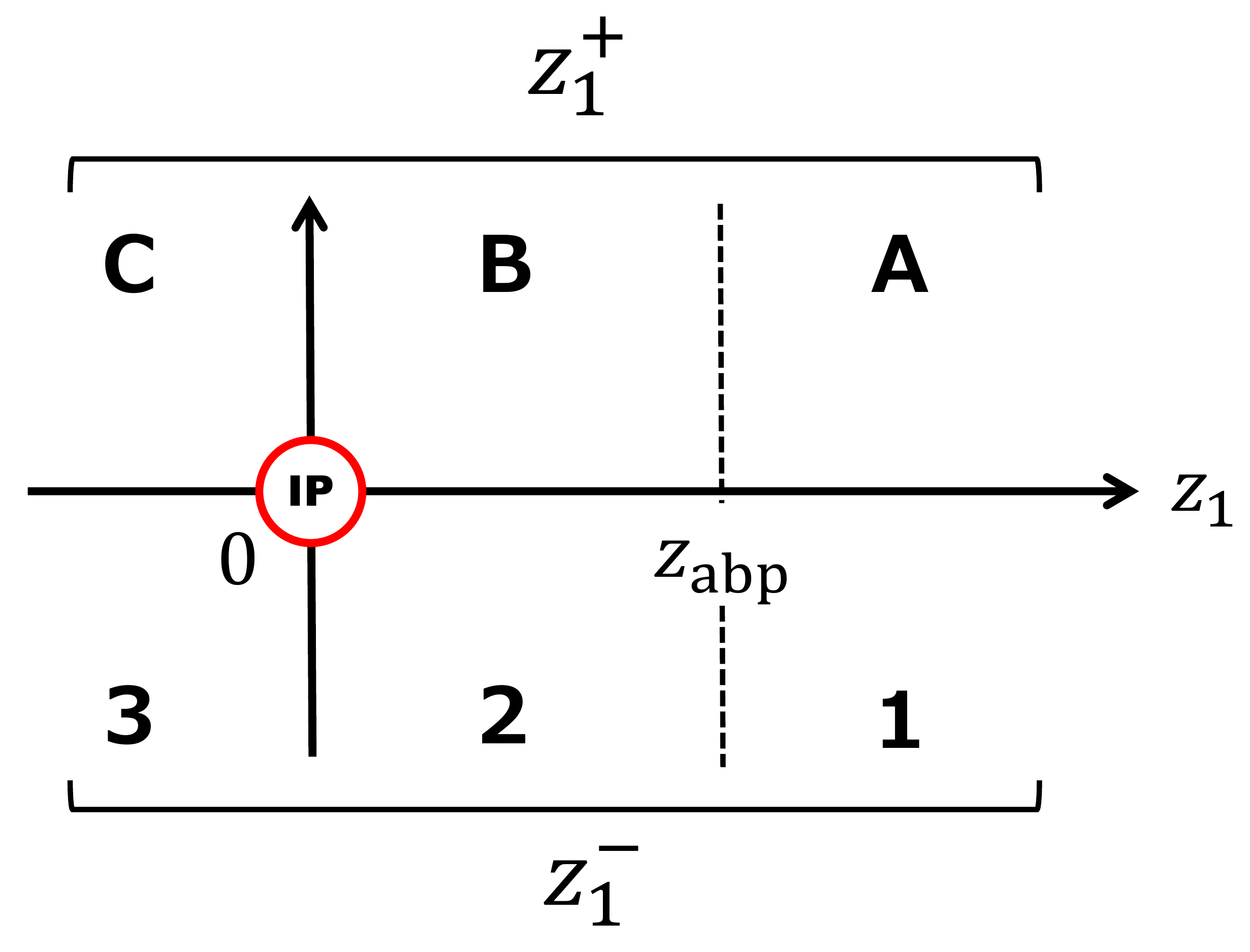}
\caption{A schematic view of the $z_1$ coordinate for the case of the scalar boson decay.}
\label{fig:z1-range2}
\end{figure}
\begin{table}[h]
\centering
\begin{tabular}{|c||c|c|c|c|}\hline
\hspace{10mm} & ~A-2~ & ~A-3~ & ~B-2~ & ~B-3~ \\\hline\hline
$z^{\rm up}$ & $z_{\rm abp}$ & $z_{\rm abp}$ & $z_1^+$ & $z_1^+$ \\\hline
$z^{\rm lw}$ & $z_1^-$        & $0$            & $z_1^-$ & $0$     \\\hline
\end{tabular}
\caption{
A summary of the integration interval of $z_1$, where $z_1^\pm$ are defined in Eqs.~(\ref{eq:z1pm}) and (\ref{eq:def-z1pm2}), while $z_{\rm abp}$ in table~\ref{tab:z0-range}.
The probabilities are zero for the other cases.
}
\label{tab:z1-range2}
\end{table}
The probability that the scalar boson $S$ decays inside the detector is give  by
\begin{align}
\mathcal{P}_S^\textrm{det} =
\frac{1}{d_X \cos\theta_0}\int_{z^{\rm lw}}^{z^{\rm up}} dz_1 ~e^{-\frac{z_1}{d_X \cos\theta_0}}
\times 
\frac{1}{d_S \cos\theta_1}\int_{L_{\rm min}-z_1}^{L_{\rm max}-z_1} dz_2 ~e^{-\frac{z_2}{d_S \cos\theta_1}}~.
\label{eq:prob_sc}
\end{align}
As is the case of the gauge boson decay, we divide the $z_1$ coordinate into three ranges as depicted in figure \ref{fig:z1-range2}.
The integral interval of $z_1$ is summarized in table~\ref{tab:z1-range2} for each case.

\bibliographystyle{apsrev}
\bibliography{biblio}

\end{document}